\documentclass[apj,twocolumn]{openjournal}
\usepackage{natbib} 
\usepackage[dvipsnames]{xcolor}
\usepackage{aas_macros} 
\usepackage{amssymb}
\usepackage{amsmath}
\usepackage[title]{appendix}
\usepackage{hyperref}	
\hypersetup{colorlinks=true,linkcolor=blue,citecolor=blue,filecolor=blue,urlcolor=blue}
\usepackage[caption=false]{subfig}
\usepackage{soul}                          

\newcommand{\angstrom}{\text{\normalfont\AA}}

\newcommand{\msunyr}{\mbox{$\rm M_{\odot}\,{\rm yr^{-1}}$}}
\newcommand{\mstar}{\mbox{$M_{*}$}}
\newcommand{\MUV}{\rm M_{UV}}
\newcommand{\mvir}{\mbox{$M_{\rm vir}$}}

\newcommand{\SFRten}{\mbox{${\rm SFR}_{\rm 10}$}}

\newcommand{\Fig}[1]{Fig.~\ref{#1}}

\newcommand{\Tab}[1]{Table~\ref{#1}}

\def\bpass  {{\sc BPASS}}
\def\adaptahop  {{\sc AdaptaHop}}
\def\cloudy  {{\sc Cloudy}}
\def\music  {{\sc MUSIC}}
\def\ramses  {{\sc Ramses}}
\def\rascas  {{\sc Rascas}}
\def\spdrone  {{\sc SPDRv1}}
\def\sphinx  {{\sc Sphinx}}
\def\sphinxtw  {{\sc Sphinx$^{20}$}}

\begin{document}
\title[The \sphinx{} Public Data Release]{The \sphinx{} public data release: Forward modelling high-redshift JWST observations with cosmological radiation hydrodynamics simulations\vspace{-15mm}}
\author{Harley Katz$^{1*}$, Joki Rosdahl$^{2}$, Taysun Kimm$^3$, Jeremy Blaizot$^{2}$, Nicholas Choustikov$^{1}$, Marion Farcy$^4$, Thibault Garel$^5$, Martin G. Haehnelt$^6$, Leo Michel-Dansac$^{2}$, and Pierre Ocvirk$^7$}
\thanks{$^*$E-mail: \href{mailto:harley.katz@physics.ox.ac.uk}{harley.katz@physics.ox.ac.uk}},

\affiliation{$^{1}$Sub-department of Astrophysics, University of Oxford, Keble Road, Oxford OX1 3RH, United Kingdom}
\affiliation{$^{2}$Centre de Recherche Astrophysique de Lyon UMR5574, Univ Lyon1, ENS de Lyon, CNRS, F-69230 Saint-Genis-Laval, France}
\affiliation{$^{3}$Department of Astronomy, Yonsei University, 50 Yonsei-ro, Seodaemun-gu, Seoul 03722, Republic of Korea}
\affiliation{$^{4}$Institute for Physics, Laboratory for Galaxy Evolution and Spectral Modelling, EPFL, Observatoire de Sauverny,
Chemin Pegasi 51, 1290 Versoix, Switzerland}
\affiliation{$^{5}$Observatoire de Genève, Université de Genève, Chemin Pegasi 51, 1290 Versoix, Switzerland}
\affiliation{$^{6}$Kavli Institute for Cosmology and Institute of Astronomy, Madingley Road, Cambridge, CB3 0HA, UK}
\affiliation{$^{7}$Observatoire Astronomique de Strasbourg, Université de Strasbourg, CNRS UMR 7550, 11 rue de l’Université, 67000 Strasbourg, France}

\begin{abstract}
The recent launch of JWST has ushered in a new era of high-redshift astronomy by providing detailed insights into the gas and stellar populations of galaxies in the epoch of reionization. Interpreting these observations and translating them into constraints on the physics of early galaxy formation is a complex challenge that requires sophisticated models of star formation and the interstellar medium (ISM) in high-redshift galaxies. To this end, we present Version 1 of the \sphinxtw{} public data release. \sphinxtw{} is a full box cosmological radiation hydrodynamics simulation that simultaneously models the large-scale process of cosmic reionization and the detailed physics of a multiphase ISM, providing a statistical sample of galaxies akin to those currently being observed by JWST. The data set contains $\sim14,000$ mock images and spectra of the stellar continuum, nebular continuum, and 52 nebular emission lines, including Ly$\alpha$, for each galaxy in \sphinxtw{} with a star formation rate $\geq0.3\ {\rm M_{\odot}\ yr^{-1}}$. All galaxy emission has been processed with dust radiative transfer and/or resonant line radiative transfer, and data is provided for ten viewing angles for each galaxy. Additionally, we provide a comprehensive set of intrinsic galaxy properties, including halo masses, stellar masses, star formation histories, and ISM characteristics (e.g., metallicity, ISM gas densities, LyC escape fractions). This paper outlines the data generation methods, presents a comparative analysis with JWST ERS and Cycle 1 observations, and addresses data set limitations. The \sphinxtw{} data release can be downloaded at the following URL: \url{https://github.com/HarleyKatz/SPHINX-20-data}.
\end{abstract}

\section{Introduction}
Elucidating the underlying physics that governs galaxy formation at high-redshift is one of the primary goals of extragalactic spectroscopic and imaging surveys with JWST \citep{Gardner2006}. However, understanding how to map the features of a galaxy image or spectra to the underlying properties of the stellar populations and gas as well as the physical processes driving the evolution of the interstellar and circumgalactic medium (ISM, CGM) represents a key theoretical challenge. 

Prior to JWST, high-redshift space-based observations were limited to primarily photometric surveys with the Hubble Space Telescope (HST) and Spitzer. The occasional follow-up observation with the WFC grism on HST or ground-based telescopes (e.g. ALMA, MOSFIRE on Keck, or MUSE on the VLT) provided additional spectroscopic redshift confirmation for some of the high-redshift candidates \citep[e.g.][]{Stark2017,Hashimoto2018,Jiang2021,Inami2017}. These telescopes have been used predominantly to constrain high-redshift population statistics such as the UV luminosity function \citep{Bouwens2015a,Livermore2017}, the global growth of star formation or UV luminosity density as a function of redshift \citep{Ellis2013,Oesch2018}, and galaxy morphology \citep[e.g.][]{Kawamata2018,Bouwens2022size}. The photometry of individual galaxies has been used to measure early star formation histories (SFHs) with spectral energy distribution (SED) fitting codes \citep[e.g.][]{Chevallard2016,Carnall2018,Johnson2021} or to infer the presence of strong emission lines (e.g. by IR excesses, \citealt{Roberts-Borsani2016}), which can be used to measure quantities such as the ionizing photon production efficiency \citep[$\xi_{\rm ion}$, e.g.][]{Bouwens2016,Stefanon2022}. JWST significantly improves upon HST+Spitzer data for numerous reasons. Not only is JWST more sensitive and has higher spatial resolution compared to its predecessors, but it also has significantly more filters that probe a wavelength range overlapping and between HST and Spitzer that is ideal for capturing rest-frame UV and optical photons at high-redshift. Hence the accuracy by which one can constrain quantities such as the SFH or stellar mass of a galaxy, purely from photometry is greatly improved with JWST. Likewise, the additional spatial resolution and sensitivity allows for spatially resolved constraints on these properties, deeper in rest-frame UV/optical luminosity that were impossible before. 

Possibly the most important advancement with JWST is its spectroscopic capabilities. Not only can photometric redshifts be readily confirmed (or not) with NIRSpec or MIRI \citep[e.g.][]{AH2023,Roberts-Borsani2023}, but emission and absorption lines can also be detected which provide an unprecedented view into the ISM and CGM of early galaxies \citep[e.g.][]{Cameron2023,Sanders2023}. The first three NIRSpec observations of $z>7$ galaxies with JWST released as part of the ERO data have already shown that the important heating mechanisms in the ISM at high-redshift are potentially significantly different compared to what is observed in the local Universe \citep{Katz2023-4363}. There are now numerous examples of high-redshift JWST galaxies where the spectrum appears nothing like the galaxies observed at low-redshift \citep[e.g.][]{Bunker2023,Isobe2023}. 

Despite the transformative nature of JWST, interpreting the observational data remains challenging. For example, SED fitting codes are the primary tools by which photometric redshifts, stellar masses, and SFHs are constrained from photometry \citep[e.g.][]{Chevallard2016,Carnall2018,Johnson2021}. Outputs from different SED fitting codes can vary significantly in their estimates of stellar population properties \citep[e.g.][]{Pacifici2023}. This is particularly problematic at high-redshift where on-going star formation can dominate the SED leaving the SFH and stellar mass estimates very sensitive to the chosen prior \citep[e.g.][]{Narayanan2023}. Furthermore, there exists significant disagreement between spatially resolved and galaxy-integrated stellar mass estimates \citep{Clara2023}.

Historically, 1D photoionization or shock models, e.g. those run with {\small CLOUDY} \citep{Ferland2017} or {\small MAPPINGS} \citep{Sutherland2017}, overwhelmingly dominate the interpretation of galaxy spectra, in particular, for measuring quantities such as the ISM electron density, temperature, metallicity, and ionization parameter. These models have been extremely successful in explaining the physics behind commonly used strong-line diagnostics at low and high redshift \citep[e.g.][]{Kewley2001,Strom2017,Nakajima2022}. Because of their 1D nature, such models are computationally inexpensive such that large grids of models with varying physical complexities can be employed to map the expected parameter space of galaxy properties from the local Universe to the epoch of reionization \citep[e.g.][]{Morisset2015}. 

However, the simplicity of photoionization and shock models also represents a limitation. The real ISM and CGM of galaxies are multiphase, exhibit turbulent geometries --- that can deviate significantly from the typically assumed gas slab or spherical cloud --- and include highly dynamic processes such as stellar winds, radiation pressure, and supernova (SN) feedback, that are much more difficult, if not impossible, to capture in 1D models. For this reason, there has been substantial recent effort to expand the capabilities of photionization models to three dimensions \citep[e.g.][]{Ercolano2008,Gray2017,Jin2022}. Moreover, in certain situations, the underlying assumptions in photoionization models can highly bias inferences of galaxy properties. This has been shown to be the case for direct method metallicity measurements, which are widely regarded  as the gold standard, but systematically underestimate metallicity in the presence of temperature fluctuations \citep[e.g.][]{Cameron2022,Delgado2023}. A similar bias also impacts electron density estimates \citep{Delgado2023b}. 

One possible solution to the challenges associated with applying SED fitting codes and photoionization models to high-redshift data is to calibrate them with low-redshift ``analog'' galaxies, for which there is ample data. This technique has been extensively applied to extreme emission line galaxies \citep[e.g.][]{Izotov2016,Amorin2017,Flury2022} to deduce properties like the Lyman-$\alpha$ (Ly$\alpha$) or Lyman continuum (LyC) escape fractions. However, there is some debate about the similarity of the low-redshift analogue population to those galaxies that form at $z\gtrsim6$ \citep[][]{Katz2023-4363,Schaerer2022}. Furthermore, the issue remains that the true underlying galaxy properties are still unknown in analogs. 

For this reason, cosmological simulations are a valuable tool for understanding galaxy formation at early times. Such simulations are important in at least two ways:
\begin{enumerate}
    \item They represent a complementary means to low-redshift analogs in testing the ability of more simplistic models to recover the true intrinsic parameters of a galaxy. Unlike the low-redshift analogs, with a cosmological simulation, all intrinsic galaxy properties are known. While all simulations are inevitably subject to subgrid modelling, the advantage of simulations is that, with sufficient resolution, they capture the complex interplay between gas accretion and cooling, star formation, and feedback in 3D, all of which can modify the morphology and spectrum of a galaxy, for example, through the star formation history, production and distribution of dust, and the thermo-chemical state of the ISM. Hence if the techniques used to infer galaxy properties do not work on simulated noise-free mock data, it is doubtful that they would generalize to real observations. Simulations are an ideal laboratory for exploring the accuracy and underlying systematic biases of SED codes and photoionization model grids.
    \item Because simulations attempt to model much of the physics of galaxy formation from first principles, comparisons between simulations and observations can be used to test our underlying theories of galaxy formation and to deduce which physical processes become important in regulating galaxy growth and evolution as a function of redshift.
\end{enumerate}

In recent years there has been a substantial increase in the number of simulations specifically targeted towards high-redshift galaxy formation. These include {\small Aurora} \citep{Pawlik2017}, {\small BlueTides} \citep{Waters2016}, {\small CoDa} \citep{Ocvirk2016}, {\small CROC} \citep{Gnedin2014}, {\small Flares} \citep{Lovell2021}, {\small Renaissance} \citep{Oshea2015}, {\small Serra} \citep{Pallottini2019}, \sphinx{} \citep{Rosdahl2018}, {\small Thesan} \citep{Kannan2022}, as well as many others. In order to 1) employ simulations as a means of evaluating the accuracy of SED fitting codes and photoionization model grids in inferring underlying galaxy physical properties and 2) to test our theories of galaxy formation by comparing simulated outputs with real observations, it is necessary to completely forward-model mock observations from simulations. While this is immediately true for the first proposed utility, for the latter it is less-obviously the case. All simulations require subgrid models that are often calibrated to reproduce some expected behaviour (see discussion in Section 2.1 of \citealt{Schaye2015}), for example, the globally inferred star formation rate density \citep{Madau2014}, the stellar mass function \citep{Li2009,Baldry2012}, the stellar mass-halo mass relation \citep{Moster2018,Behroozi2019}, the mass-metallicity relation \citep{Tremonti2004}, or the UV luminosity function \citep{Bouwens2015a}. The crucial aspect of these calibrations is that the target quantities are often inferred from observations rather than being the directly observed. 

Strong systematic biases often exist between simulations and these observational inferences. For example, the calibration between star formation rate (SFR) and H$\alpha$ luminosity is sensitive to both the chosen stellar initial mass function (IMF), underlying single stellar population (SSP) model, and metallicity \citep[e.g.][]{Kennicutt2012,Eldridge2017}. Stellar masses are subject to the prior on SFR \citep[e.g.][]{Narayanan2023}, while metallicity is highly method-dependent \citep{Andrews2013} and often biased towards that of H~{\small II regions}. Something as straightforward as a half-light radius is subject to how the image is segmented (which sets the outer radius), whether the quantity is circularized, if a Sérsic fit \citep{Sersic1963} is used, whether the Sérsic fit is done in 1D or 2D, and whether the slope of the Sérsic profile is fixed to a specific value (e.g. one). These definitions not only differ between simulations and observers, but the method used to report the same quantity in different observational papers often differs. In order to be able to apply the same measurement techniques to both simulated and observed data and make valid comparisons, it is paramount that simulations be forward modelled into observational space. 

To forward model the stellar continuum, dust radiative transfer is a common technique \citep[e.g.][]{Jonsson2006,Baes2011}, and this has widely been applied to large simulations \citep[e.g.][]{Camps2018,Feldmann2023,Yi2023}. However, at high-redshift, nebular emission becomes an important part of a galaxy SED \citep[e.g.][]{Schaerer2010}. While SSP libraries are often shipped with emission line estimates \citep[e.g.][]{Byler2017,Xiao2018}, the underlying assumptions of these models (e.g. gas density, ionization parameter) are not obviously representative of the physics of the high-redshift ISM. Hence, simply applying these to simulated star particles may have limited fidelity for high-redshift galaxies. A more accurate technique would be to try to simulate the high-redshift ISM and its ionization state from first principles. This is particularly important for simulations that model resonant emission lines such as Ly$\alpha$ \citep{Blaizot2023} or low ionization state absorption features \citep{Gazagnes2023}. While radiative transfer is now included in many simulations \citep[e.g.][]{Pawlik2017,Kannan2022,Ocvirk2016}, the vast majority of full-box radiative hydrodynamics simulations lack a model for the ISM. 

\sphinx{} \citep{Rosdahl2018} and its successor \sphinx$^{20}$ \citep{Rosdahl2022} are notable exceptions as they simulate a statistically large sample of galaxies in a full cosmological volume with radiative hydrodynamics while simultaneously reaching resolutions high enough ($\approx10$ physical pc) to begin resolving and modelling the physics of a multiphase ISM. Moreover, the volume of the simulation is large enough such that many of the galaxies are bright enough to be observed with telescopes such as JWST, ALMA, and HST. Hence, \sphinx{} is a unique resource where the full SEDs, including stellar continuum, nebular continuum, and nebular emission lines, can be predicted in a spatially-resolved manner and to a high-degree of accuracy for a large sample of high-redshift galaxies. This mock data can be used to directly compare with JWST data and as a laboratory for inferring galaxy properties from mock observations. 

In this work, we present Version 1 of the \sphinx{} Public Data Release (\spdrone). The data set is an open-sourced library of $\sim14,000$ mock images and spectra for galaxies in the redshift range $4.64\leq z\leq 10$ with 10 Myr-averaged star formation rates $\geq0.3\ {\rm M_{\odot}\ yr^{-1}}$, specifically designed to be compared with JWST observations. We have supplemented the data set with relevant intrinsic galaxy properties and made the repository publicly accessible at the following URL: \url{https://github.com/HarleyKatz/SPHINX-20-data}. Below we describe the methods used to produce \spdrone\ and give a tour of what is available in the data set by comparing the results with some of the early ERS and Cycle~1 JWST observations.

\section{Methods}
In order to make direct comparisons between \sphinx{} and spectroscopic and photometric observations of galaxies in the early Universe, we must forward model both the intrinsic emission of the gas and stars in the simulated galaxies and the propagation of these photons through the intervening ISM and CGM. We consider both stellar and nebular (lines and continuum) emission as well as attenuation and scattering by dust and/or H~{\small I}. Below we describe the details for generating each component of the mock spectra and images for each \sphinx{} galaxy.

\subsection{The \sphinx{}$^{20}$ Simulations}

\sphinx{} is a suite of cosmological radiation-hydrodynamic simulations with state-of-the-art physical models and spatial/mass resolution designed to simultaneously capture a multiphase ISM on the scale of galaxies, and the reionization of the intergalactic medium (IGM) on large scales. The simulations have been used to discern the physics of early galaxy formation and the process of reionization, illustrating the sensitivity of the escape of LyC photons from galaxies to stellar evolution models \citep{Rosdahl2018}, the creation/amplification and role of magnetic fields \citep{Attia2021, Katz2021}, how reionization suppresses the growth of dwarf galaxies \citep{Katz2020reion} and drives the observability of Lyman-alpha emitters at $z\gtrsim6$ \citep{Garel2021}, the redshift-evolution of metal-line emission \citep{Katz2022-CII-OIII}, how the LyC escape from galaxies correlates with observable galaxy properties \citep{Maji2022, Katz2022-mgii, Choustikov2023}, and which galaxies predominantly power reionization \citep{Rosdahl2022, Katz2023-CII}.

\sphinxtw{}, which is exclusively used in this data release, is the largest volume simulation in the \sphinx{} suite, with a volume of 20$^3$ co-moving Mpc$^3$. It contains thousands of star-forming galaxies at each snapshot between $z=4.64$ and $z=10$.
Here, we summarize the most relevant characteristics of the simulation, but refer to \cite{Rosdahl2022} and \cite{Rosdahl2018} for full details.

\sphinxtw{} was run over the period from February~2019 to November~2020 on the JUWELS\footnote{\url{https://www.fz-juelich.de/en/ias/jsc/systems/supercomputers/juwels}} and Joliot-Curie ROME\footnote{\url{https://www-hpc.cea.fr/en/Joliot-Curie.html}} supercomputers and used 63 million CPU-hours to reach a final redshift of 4.64, when the volume is completely reionized. As for all \sphinx{} volumes, we use the \ramses{} adaptive mesh refinement code \citep{Teyssier2002} with full radiation-hydrodynamics to model the propagation of LyC photons and their interactions with hydrogen and helium \citep{Rosdahl2018, Rosdahl2015}. Gas cells are refined on mass and Jeans length, with a minimum cell width of 76 co-moving pc, corresponding to 11 physical pc at $z=6$. The dark matter particle mass is $2.5 \times 10^5~{\rm M_{\odot}}$, allowing halos to be resolved approximately down to the atomic cooling limit, and stellar particles, representing stellar populations, have initial masses of $400~{\rm M_{\odot}}$. Stars are formed out of the gas with a variable local efficiency that depends on the thermo-turbulent properties of the gas. Supernova feedback is performed using the mechanical model described in \cite{Kimm2014,Kimm2015}, that adapts to the local conditions in order to generate a consistent final momentum from the Sedov-Taylor phase. To obtain strong enough suppression in star formation, we assume 4 SN explosions per 100~$~{\rm M_{\odot}}$, which is significantly more than found with standard initial mass functions observed in our local environment \citep[but see][]{Katz2022-CII-OIII}. \sphinx{} does not include black hole formation or AGN feedback due to its small volume, although new data is suggesting that black holes may be more relevant than realised at early epochs \citep{Greene2023,Matthee2023b,Kocevski2023}. Gas and stellar metallicity is tracked with a single scalar, and we describe below how we extract individual metal species and ionization states of the gas in post-processing. We use the \bpass{} version 2.2.1 SED model \citep{Stanway2018} to calculate  the injected LyC luminosities of stellar populations as a function of their ages and metallicities. We use two LyC radiation groups on-the-fly (13.6~eV-24.59~eV and 24.59~eV-$\infty$), but post-process the simulation snapshots used in this data release with two additional lower-energy radiation groups (5.6~eV-11.2~eV and 11.2~eV-13.6~eV) to more accurately predict metal ionization states with ionization potentials lower than H~{\small I}. Non-equilibrium thermochemistry of hydrogen and helium is performed on-the-fly, with the contribution of metals to cooling tabulated using \cloudy{} for high temperatures and approximated with the fit from \cite{Rosen1995} for low temperature. The cosmological initial conditions are generated using \music{} \citep{Hahn2011} and have been selected to produce a typical patch of the Universe in terms of the halo mass function. We use $\Lambda$CDM cosmological parameters compatible with the \cite{Ade2014}: $\Omega_{\rm m} = 0.3175$, $\Omega_{\Lambda} = 0.6825$, $\Omega_{\rm b} = 0.049$, $\sigma_8 = 0.83$, $H_0 = 67.11 \ \ {\rm km\ s^{-1}\ Mpc^{-1}}$.

\adaptahop{} \citep{Tweed2009} is used for dark matter halo identification, with halo finder parameters as described in \cite{Rosdahl2022}. For each halo, we designate its occupant galaxy to consist of all gas and stellar particles enclosed within its virial radius.

In order to maintain manageable data sizes, we limit the current data release to seven \sphinxtw{} snapshots, spanning  $z=10$ to $z=4.64$, and only galaxies which have star formation rates larger than $0.3 \ \msunyr$, i.e. omitting the underlying population of galaxies likely too dim to be observed with current state-of-the-art telescopes. Furthermore, we only consider main haloes as all subhaloes will be contained within the virial radius. We list in \Tab{tab:sample} the snapshots and the number of selected galaxies in each, totaling to 1,380 galaxies.

\begin{table}
 \caption{Statistics of the galaxies in the \sphinx{} data release. Each column shows the number of galaxies with ${\rm SFR}\geq0.3\,\msunyr$, their maximum and median stellar mass, and the maximum and median virial mass of their host dark matter halo, respectively.}
 \label{tab:sample}
 \centering
 \begin{tabular}{cccccc}
  \hline
  Redshift & $N_{\rm gal}$ & $\log \mstar$  & $\log \mstar$ &  $\log M_{\rm vir}$ &  $\log M_{\rm vir}$ \\
  &  & Max & Median & Max & Median\\
  \hline
 10.0 &  49  &  8.56 &  7.38 & 10.20 &  9.30 \\
 9.0  &  66  &  8.72 &  7.52 & 10.48 &  9.44 \\
 8.0  & 128  &  9.25 &  7.62 & 10.75 &  9.55 \\
 7.0  & 177  &  9.65 &  7.87 & 10.91 &  9.69 \\
 6.0  & 276  &  9.93 &  8.07 & 11.12 &  9.89 \\
 5.0  & 317  & 10.46 &  8.27 & 11.64 & 10.00 \\
 4.6  & 367  & 10.63 &  8.40 & 11.70 & 10.07 \\
  \hline
 \end{tabular}
\end{table}

\subsection{Stellar continuum generation}
The intrinsic stellar continuum is generated for each halo by summing the contributions of each star particle inside the virial radius of the halo. For each star particle, we assign a spectrum based on its age, mass, and metallicity by interpolating the model SEDs from BPASS v2.2.1 \citep{Stanway2018}. We assume the same upper mass cutoff (100~${\rm M_{\odot}}$) and IMF slope ($-1.35$) as was used to calculate the ionizing luminosities for star particles in the simulations.

\subsection{Emission line generation}
Intrinsic emission line luminosities are calculated for 52 emission lines (see Table~\ref{tab:emission_lines}) for each gas cell within the virial radius of the halo following a modification to the method described in \cite{Katz2022-mgii} as detailed in \cite{Choustikov2023}. We first perform a search for all gas cells that host star particles where the Stromgren sphere is unresolved (i.e. $R_{\rm S}<\Delta x/2$). For these cells, we calculate emission line luminosities by interpolating tables of {\small CLOUDY} models that were generated with v17.03 \citep{Ferland2017}. The table contains models varying stellar age (computed as the ionizing luminosity-weighted mean of all star particles within the gas cell), metallicity, gas density, and total ionizing luminosity. These models assume a spherical geometry (since the H~{\small II} region is fully embedded within the host gas cell). Because \sphinx{} does not follow the enrichment of individual elements, metal abundances are assumed to follow \cite{Grevesse2010} scaled by the metallicity of the cell. The models are iterated to convergence and stopped at an electron fraction of 1\%. For all other gas cells we assume that the radiation and electrons are well mixed. We tabulate line emissivities from a grid of one-zone slab models (again using {\small CLOUDY}) varying the gas density, metallicity, ionization parameter, and electron fraction. All models are iterated to convergence and the shape of the SED varies with metallicity, but is assumed to have an age of $10$~Myr. Furthermore, for cells with resolved ionization fractions, emission lines coming from H and He use the non-equilibrium ionization fractions directly from the simulation to compute luminosities. The total intrinsic emission of each halo is then the sum of of the intrinsic line luminosities of all cells within the virial radius.

\begin{table}
    \centering
    \caption{Emission lines included in the \sphinx{} data release. Species are defined by their element symbol and ionization state. States with a suffix of ``R'' or ``C'' represent the recombination or charge exchange contribution to a collisionally excited line. Emission lines shown in magenta have been propagated through dust and/or H~I radiative transfer. Wavelengths are the default values in cloudy v17.}
    \centering
    \begin{tabular}{lcr|lcr}
    \hline
    Species & State & Wavelength & Species & State & Wavelength  \\
    \hline
    \textcolor{magenta}{H}  & \textcolor{magenta}{1} & \textcolor{magenta}{1215.67 \AA}  & O  & 3 & 51.80 $\mu$m \\
    \textcolor{magenta}{H}  & \textcolor{magenta}{1} & \textcolor{magenta}{6562.80 \AA}  & O  & 3 & 88.33 $\mu$m \\
    \textcolor{magenta}{H}  & \textcolor{magenta}{1} & \textcolor{magenta}{4861.32 \AA}  & \textcolor{magenta}{Ne} & \textcolor{magenta}{3} & \textcolor{magenta}{3868.76 \AA} \\
    \textcolor{magenta}{H}  & \textcolor{magenta}{1} & \textcolor{magenta}{4340.46 \AA}  & Ne & 3 & 3967.47 \AA  \\
    \textcolor{magenta}{H}  & \textcolor{magenta}{1} & \textcolor{magenta}{4101.73 \AA}  & C  & 2 & 157.64 $\mu$m \\
    He & 2 & 1640.41 \AA  & C  & 3 & 1906.68 \AA  \\
    He & 2 & 4685.68 \AA  & C  & 3 & 1908.73 \AA  \\
    O  & 1 & 6300.30 \AA  & C  & 4 & 1548.19 \AA  \\
    O  & 1 & 6363.78 \AA  & C  & 4 & 1550.78 \AA  \\
    \textcolor{magenta}{O}  & \textcolor{magenta}{2} & \textcolor{magenta}{3726.03 \AA}  & N  & 2 & 5754.61 \AA  \\
    \textcolor{magenta}{O}  & \textcolor{magenta}{2} & \textcolor{magenta}{3728.81 \AA}  & N  & 2R & 5755.00 \AA  \\
    O  & 2R & 3726.00 \AA  & N  & 2 & 6548.05 \AA  \\
    O  & 2R & 3729.00 \AA  & \textcolor{magenta}{N}  & \textcolor{magenta}{2} & \textcolor{magenta}{6583.45 \AA}  \\
    O  & 2 & 7318.92 \AA  & N  & 2R & 6584.00 \AA  \\
    O  & 2 & 7319.99 \AA  & N  & 3 & 1748.65 \AA  \\
    O  & 2 & 7329.67 \AA  & N  & 3 & 1753.99 \AA  \\
    O  & 2 & 7330.73 \AA  & N  & 3 & 1746.82 \AA  \\
    O  & 2R & 7332.00 \AA  & N  & 3 & 1752.16 \AA  \\
    O  & 2R & 7323.00 \AA  & N  & 3 & 1749.67 \AA  \\
    O  & 3 & 1660.81 \AA  & S  & 2 & 6716.44 \AA  \\
    O  & 3 & 1666.15 \AA  & S  & 2 & 6730.82 \AA  \\
    \textcolor{magenta}{O}  & \textcolor{magenta}{3} & \textcolor{magenta}{4363.21 \AA}  & S  & 2 & 4076.35 \AA  \\
    O  & 3R & 4363.00 \AA  & S  & 2 & 4068.60 \AA  \\
    O  & 3C & 4363.00 \AA  & S  & 3 & 6312.06 \AA  \\
    \textcolor{magenta}{O}  & \textcolor{magenta}{3} & \textcolor{magenta}{4958.91 \AA}  & S  & 3 & 9068.62 \AA  \\
    \textcolor{magenta}{O}  & \textcolor{magenta}{3} & \textcolor{magenta}{5006.84 \AA}  & S  & 3 & 9530.62 \AA  \\
    \hline 
    \end{tabular}
    \label{tab:emission_lines}
\end{table}

\subsection{Nebular continuum generation}
The final component of the spectrum is the nebular continuum. Here we again split cells based on whether or not they host an unresolved Stromgren sphere. In the case where there is an unresolved Stromgren sphere, we again interpolate tabulated {\small CLOUDY} data for the nebular continuum (free-free, free-bound, two-photon) of only H and He. For all other cells we compute the free-free, free-bound, and two-photon emission for H and He individually using the non-equilibrium ionization fractions and electron densities in the cell. Free-free emission is computed following the method described in \cite{Schirmer2016} using Gaunt factors from \cite{vanhoof2014}. Free-bound emission is computed with {\small CHIANTI} \citep{Dere2019}. Finally, two-photon emission is also computed following \cite{Schirmer2016}.

\subsection{Dust and Resonant Line Radiative Transfer}
All line and continuum radiation is subject to absorption and scattering by dust. The \sphinx{} public data release also contains bright resonant lines such as Ly$\alpha$ that must diffuse both spatially and in frequency space to escape the galaxy and be viewed by an observer. All of this physics is modelled using the Monte-Carlo radiative transfer code \rascas{} \citep{Michel-Dansac2020} similarly to the method described in \cite{Katz2022-mgii}. Dust is assigned to all gas cells using the Small Magellanic Cloud (SMC) dust model described in \cite{Laursen_2009}. This phenomenological dust model is normalized such that the dust absorption coefficient in each gas cell is given by $(n_{\rm HI}+f_{\rm ion}n_{\rm HII})\sigma_{\rm dust}Z/Z_0$, where $n_{\rm HI}$ and $n_{\rm HII}$ are the number densities of neutral and ionized hydrogen, respectively, the fraction of ionized gas contributing to the dust density is assumed to be $f_{\rm ion}=0.01$, $\sigma_{\rm dust}$ is the dust cross section, $Z$ is the metal mass fraction, and $Z_0=0.005$. The dust asymmetry and albedo are adopted from the SMC dust model of \cite{Weingartner2001}.

For the emission lines shown in magenta in Table~\ref{tab:emission_lines}, 200,000 photon packets are probabilistically distributed to all gas cells of a halo based on their intrinsic luminosities. The initial wavelengths of the photon packets are placed at line-centre of the transition, thermally broadened, and shifted to the observer's frame based on the bulk velocity of the cell. Escape fractions and images for all other emission lines are interpolated from the closest ion for which we run the radiative transfer due to computational constraints. For the nebular and stellar continuum, we sample the spectrum at 20 different wavelengths between 1300~\AA~and 6583~\AA. The wavelengths are chosen to be consistent with the locations of strong emission lines (so that equivalent widths can be accurately measured) and consistently throughout the UV where dust attenuation is the strongest. Another 200,000 photon packets are probabilistically distributed to each star particle based on the luminosity of the stellar continuum at each wavelength. Likewise, the same number of packets are probabilistically distributed for each wavelength to gas cells based on the nebular continuum luminosity. The full spectrum of each galaxy can thus be constructed by computing the escape fractions of each component and interpolating across wavelength. 

Due to the H~{\small I} opacity, Ly$\alpha$ is computed slightly differently to the method described above. Ly$\alpha$ photon packets are probabilistically assigned to gas cells based on their intrinsic luminosity, which is the sum of the recombination and collisional components. We use $10^6$ photon packets to better capture the spectrum when the escape fraction is low. The Ly$\alpha$ photon packets are propagated in a medium that consists of H~{\small I}, D, and dust. The dust model is the same as used for all of our other {\small RASCAS} runs as described above and we assume a fixed D/H abundance of $3\times10^{-5}$.

To produce mock observations along individual sight lines, we employ the peeling algorithm \citep{Yusuf_Zadeh_1984,Zheng_2002,Dijkstra_2017} to make images and spectra along ten directions. The viewing angles are kept fixed for every galaxy. As our data set contains 1,380 galaxies, viewing each along ten sight lines results in 13,800 total spectra (not including the 1,380 intrinsic spectra). 

\subsection{Available Data}
In Table~\ref{tab:galprops} we list the full set of galaxy properties, mock observations, and derived quantities that can be downloaded from the public repository. The data set contains numerous intrinsic properties of the galaxies such as stellar mass, halo mass, SFR averaged over multiple time scales, SFHs, etc. as well as the full spectra described above. We additionally provide certain derived quantities such as UV continuum slopes, UV magnitudes, and JWST filter magnitudes. The breadth of data should encapsulate the vast majority of JWST observations available at present.

\begin{table*}
 \caption{Details of the galaxy properties available as part of the \sphinx{} public data release.}
 \label{tab:galprops}
 \begin{tabular}{p{0.2\textwidth}p{0.15\textwidth}p{0.55\textwidth}}
  \hline
  Quantity & Units & Notes \\
  \hline
  Halo ID & & \\
  
  Redshift & & \\
  
  Halo mass & ${\rm \log_{10}(M/M_{\odot})}$ & \\
  
  Stellar mass & ${\rm \log_{10}(M/M_{\odot})}$ & This value is the total stellar mass formed (i.e. the integral of the star formation history) and is not adjusted for mass loss due to stellar feedback \\

  $R_{\rm vir}$ & 20~cMpc & \\

  $x,\ y,\ z$ position & 20~cMpc & 3D position of the halo in the simulation volume\\

  Star formation rate & ${\rm M_{\odot}\ yr^{-1}}$ & Provided as an average over 3, 5, 10, and 100~Myr and can be recomputed for any other interval from the star formation history \\

  Star formation history & ${\rm M_{\odot}\ yr^{-1}}$ & Provided for every galaxy on a 1~Myr time cadence \\

  Stellar ages & Myr & Mass-weighted and LyC luminosity-weighted \\

  Stellar metallicity & Absolute & Mass-weighted and LyC luminosity-weighted over all stars \\

  Stellar metallicity history & Absolute & Mass-weighted stellar metallicity of all star particles that formed in bins of 1~Myr \\

  Ionizing luminosity & photons ${\rm s^{-1}}$ & \\

  LyC escape fraction & & Angle-averaged (for all photons with $E>13.6\ {\rm eV}$) and along ten sight lines (for photons with a wavelength of 900~\AA) \\

  ISM gas density & $\log_{10}({\rm n_H/cm^{-3}})$ & Weighted by intrinsic [O~{\small II}]~$\lambda\lambda$3727 or [C~{\small III}]~$\lambda\lambda$1908 \\

  Gas metallicity & $\log_{10}(Z/Z_{\odot})$ & Mass-weighted as well as [O~{\small II}]~$\lambda\lambda$3727, [O~{\small III}]~$\lambda$5007, [N~{\small II}]~$\lambda$6583, and H$\beta$ weighted \\
  
  Emission line luminosities & ${\rm erg\ s^{-1}}$ & Intrinsic for all emission lines listed in Table~\ref{tab:emission_lines}, dust attenuated along ten sight lines for H$\alpha$, H$\beta$, H$\gamma$, H$\delta$, [O~{\small II}]~$\lambda\lambda$3727, [Ne~{\small III}]~$\lambda$3869, [O~{\small III}]~$\lambda$4363, [O~{\small III}]~$\lambda$4959, [O~{\small III}]~$\lambda$5007, [N~{\small II}]~$\lambda$6583\\

  Stellar continuum luminosities & ${\rm erg\ s^{-1}\ \angstrom^{-1}}$ & Intrinsic \& dust attenuated along ten sight lines for 20 wavelengths (1300\AA, 1400\AA, 1500\AA, 1600\AA, 1700\AA, 1800\AA, 1900\AA, 2000\AA, 2500\AA, 3000\AA, 3727\AA, 3869\AA, 4102\AA, 4341\AA, 4363\AA, 4861\AA, 4959\AA, 5008\AA, 6563\AA, 6583\AA) \\

  Nebular continuum luminosities & ${\rm erg\ s^{-1}\ \angstrom^{-1}}$ & Intrinsic \& dust attenuated along ten sight lines for 20 wavelengths (1300\AA, 1400\AA, 1500\AA, 1600\AA, 1700\AA, 1800\AA, 1900\AA, 2000\AA, 2500\AA, 3000\AA, 3727\AA, 3869\AA, 4102\AA, 4341\AA, 4363\AA, 4861\AA, 4959\AA, 5008\AA, 6563\AA, 6583\AA) \\

  Full SEDs & ${\rm erg\ s^{-1}\ Hz^{-1}\ cm^{-2}}$ & Intrinsic \& dust attenuated along ten sight lines and redshifted to the relevant $z$. Spectra are computed at 1~\AA~resolution by interpolating the escape fractions at the 20 continuum wavelengths and for each emission line. SED files provide the total SED as well as the three separate components \\

  E(B$-$V) & & Along ten sight lines. Computed from the Balmer decrement (H$\alpha$ and H$\beta$) \\ 

  Effective radii ($R_{\rm eff}$) & pc & Measured in each of the JWST filters along each line of sight for the largest segment after our image segmentation procedure. We provide the corresponding flux density (nJy) of the segment in addition to its circularized radius \\

  UV continuum slopes ($\beta$) & & Intrinsic \& dust attenuated along ten sight lines. Measured from the full SED (stellar + nebular continuum) as well as only the stellar continuum. Additional values can be measured from the photometry with the inclusion of emission lines \\

  UV magnitudes & AB & Intrinsic \& dust attenuated along ten sight lines. Measured at 1500~\AA~from the stellar and nebular continuum \\

  JWST filter magnitudes & AB & Dust attenuated along ten sight lines. Computed for all NIRCam wide and medium filters (F070W, F090W, F115W, F140M, F150W, F162M, F182M, F200W, F210M, F250M, F277W, F300M, F335M, F356W, F360M, F410M, F430M, F444W, F460M, F480M) \\

  Ly$\alpha$ and H$\alpha$ spectra & ${\rm erg\ s^{-1}}$ & Dust attenuated along ten sight lines. Spectral resolution of 0.1~\AA. Values should be divided by the wavelength bins to obtain appropriate units \\

  Ly$\alpha$ and H$\alpha$ surface brightness profiles & ${\rm erg\ s^{-1}}$ & Dust attenuated along ten sight lines. Spatial resolution of $R_{\rm vir}/250$. Values should be divided by the pixel size to obtain surface brightness \\

  Galaxy images & ${\rm nJy\ pixel}$ & Dust attenuated along ten sight lines for each JWST filter. Due to data size, these are made available upon request for any emission line or continuum (nebular or stellar) wavelength. Example RGB images combining multiple filters are shown in Figure~\ref{fig:hero_img}\\
  
  \hline
 \end{tabular}
\end{table*}

\section{Results}
The primary purpose of \spdrone\ is to help better understand the physical properties of high-redshift galaxies that are (potentially) observable with existing facilities. Thus, we have limited our analysis to actively star-forming galaxies with a 10~Myr-averaged SFR $\geq 0.3\,\msunyr$. Below we provide a tour of data set, organized by:
\begin{itemize}
    \item Intrinsic galaxy properties (e.g. stellar and halo masses, metallicities, etc.),
    \item Photometric properties (i.e. those that can be compared to observational data sets with imaging), and
    \item Spectroscopic properties (e.g. full SEDs and emission lines).
\end{itemize}
Our aim is to demonstrate the diversity of what is available as part of the data release as well as what can be computed. We provide a few sample comparisons with real observations to contextualize the data release among state-of-the-art observational data. 

\begin{figure*}
\centering
\includegraphics[width=\textwidth]{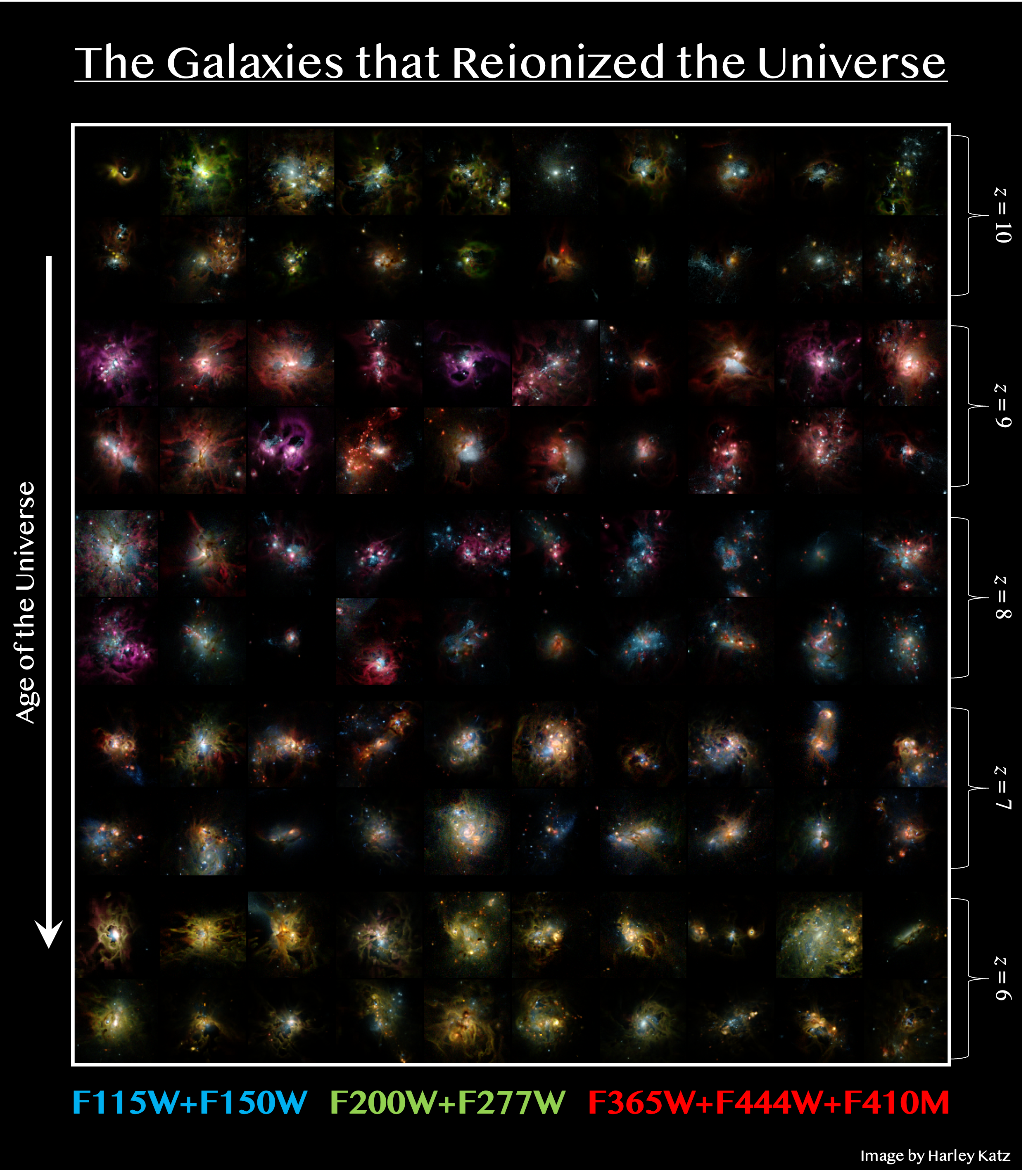}
\caption{RGB images of the \sphinx{} galaxies with the highest escaping LyC luminosity (subject to a stellar mass threshold) at each redshift. F365W, F444W, and F410M are used for the red channel, F200W, and F277W are used for green, and F115W and F150W are used for blue, identical to the UNCOVER mosaic \protect\citep{Bezanson2022}. As redshift changes, the filters sample a different wavelength range of the SEDs of the galaxies and hence they appear to change colour.}
\label{fig:hero_img}
\end{figure*}

To begin, in Figure~\ref{fig:hero_img} we show a gallery of mock RGB images of 100 \sphinx{} galaxies in the redshift interval $6\leq z\leq10$. Following the UNCOVER mosaic \protect\citep{Bezanson2022}, we combine F365W, F444W, and F410M in the red channel, F200W, and F277W in green, and F115W and F150W in blue. To highlight the lower surface brightness features, we scale the pixel intensities by an arcsinh filter. Galaxies are selected as those with the highest escaping LyC luminosities (i.e. the galaxies that are currently reionizing the simulated volume) subject to a stellar mass threshold. As redshift changes, the filters sample a different wavelength range of the SEDs of the galaxies and hence they appear to change colour. The morphological diversity exemplifies how the galaxies leaking LyC radiation can look very different. This is true not only between different galaxies, but also for the same galaxy along different lines of sight. However, in the vast majority of galaxies, the ISM is clearly disrupted and there are strong outflows, allowing for the escape of LyC radiation.  

\begin{figure*}
\includegraphics[]{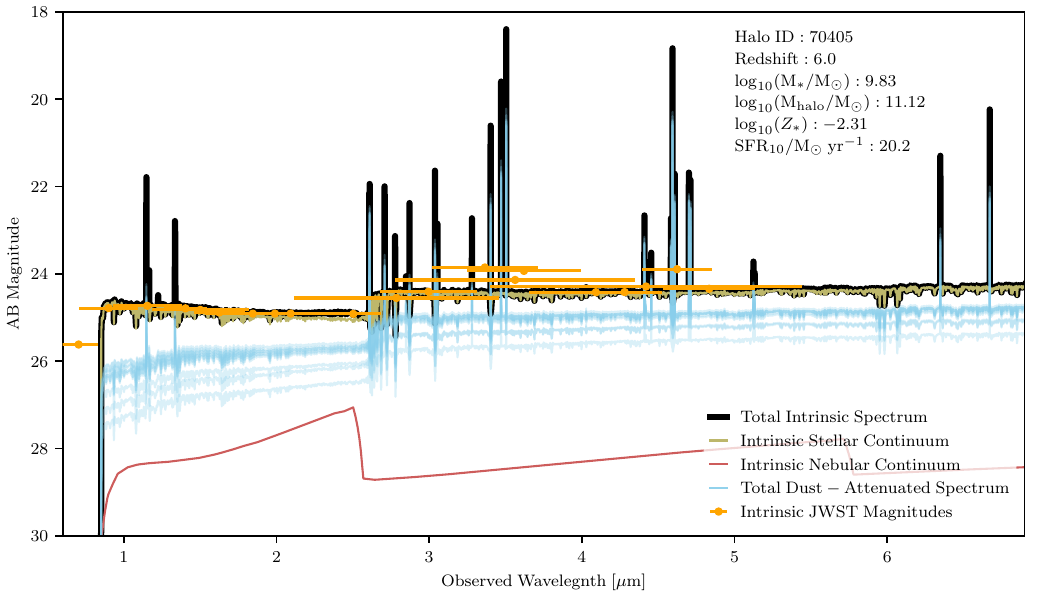}
\caption{Example spectrum of a massive $z=6$ galaxy in the \sphinx{} public data release. We show the intrinsic stellar continuum (olive), intrinsic nebular continuum (red), and total intrinsic spectrum (black). NIRCam filter magnitudes are shown as the orange data points. The different blue lines represent the dust-attenuated spectra along the ten different sight lines in the data set.}
\label{fig:spec-example}
\end{figure*}

In Figure~\ref{fig:spec-example} we show the full intrinsic spectrum (black) broken down into stellar continuum, nebular continuum, and nebular emission lines, as well as the dust-attenuated spectra along the ten sight lines for the most massive $z=6$ galaxy. Data points in this image represent the JWST filter magnitudes for the 20 wide and medium filters on NIRCam. Data such as that shown in Figures~\ref{fig:hero_img} and \ref{fig:spec-example} is available for every galaxy in \spdrone.

 \begin{figure*}
 \centering
 \includegraphics[width=0.45\textwidth]{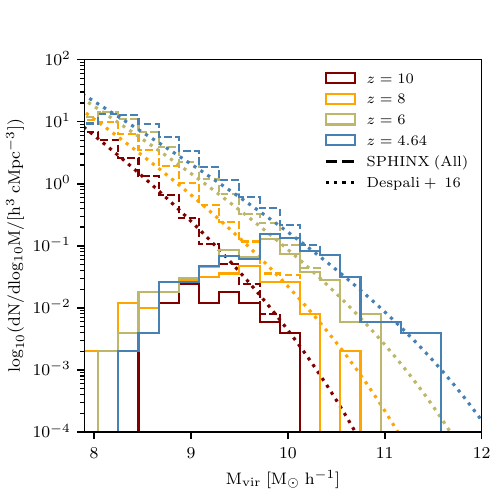}
 \includegraphics[width=0.45\textwidth]{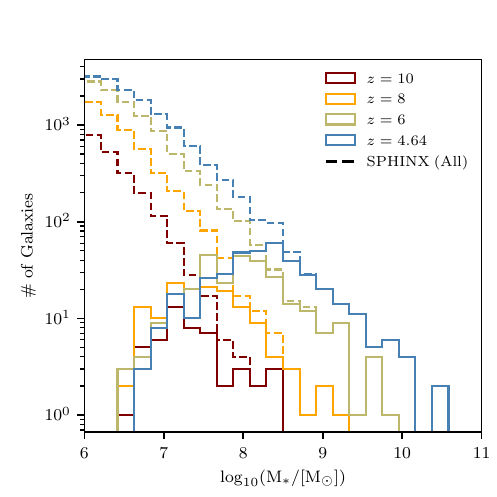}
      \caption{(Left) Dark matter halo mass functions for haloes that host star-forming galaxies with $\rm SFR_{\rm 10}\geq0.3\,M_{\odot}/yr $ in \sphinx{} for different redshifts (solid lines). For comparison, the mass functions for the entire sample of main halo population are shown as dashed lines, while the theoretical estimates by \citet{Despali2016} are shown as dotted lines. (Right) Stellar mass distributions of our main sample with $\rm SFR_{\rm 10}\geq 0.3\,M_\odot/yr$ for different redshifts (solid lines). The stellar mass distributions for all \sphinx{} galaxies are shown as dashed lines. Note that ${\rm M_{*}}$ corresponds to the total stellar mass formed, so does not take into account the mass loss due to stellar evolution.}
      \label{fig:hmf}
  \end{figure*}

\subsection{Intrinsic Galaxy Properties}
\subsubsection{Halo masses}
The initial conditions for \sphinx{} were chosen to simulate an average region of the Universe in which the total ionizing luminosity is closest to its mean value \citep[e.g., see Figure~1 of][]{Rosdahl2018}. The mass function of dark matter halos in \sphinxtw{} is shown in Figure~\ref{fig:hmf}, at different redshifts in the interval $4.6<z<10$. It is in good agreement with the theoretical expectations from dark matter-only N-body simulations \citep[e.g.,][]{Despali2016} despite the limited volume of the simulation.

\sphinx{} has 278, 1,069, 2,789, 4,508 DM halos with masses above $10^{9}~{\rm M_{\odot}}$ at $z=10$, $8$, $6$, and $4.6$, respectively. Our data release includes galaxies with $\SFRten\geq 0.3\,\msunyr$ that are mostly embedded in halos with $\mvir\sim 10^{9}~{\rm M_{\odot}}$ at $z=10$ or halos with $\mvir\sim 10^{10}~{\rm M_{\odot}}$ at $z=4.6$, but a small fraction of them is also found in halos with $10^{8}<\mvir<10^{9}~{\rm M_{\odot}}$, as star formation can be bursty and temporarily enhanced. This effect is important as analytic models often assume that SFR scales with the accretion rate of dark matter and the scatter may be important for explaining the number of bright high-redshift galaxies \citep[e.g.][]{Shen2023,Sun2023}. 

\subsubsection{Stellar masses}

In the right panel of Figure~\ref{fig:hmf}, we show the distributions of stellar masses formed (${\rm M_{*,i}}$) in \sphinxtw galaxies. ${\rm M_{*,i}}$ is computed as the sum of the initial mass of all star particles within the virial radius (i.e. including subhaloes). This is equivalent to computing the integral of the SFH of each galaxy and neglects mass loss due to stellar evolution. Because we consider all star particles within the virial radius, the data release excludes subhaloes to avoid double counting.

Within the $20^3 \,{\rm cMpc}^3$ volume, \sphinxtw{} has 2,417, 6,099, 11,747, and 15,455 galaxies with ${\rm M_{\rm *,i}>10^6\ M_{\odot}}$ at $z=10$, $8$, $6$, and $4.64$, respectively. Of these, only 49, 128, 276, and 367 galaxies with ${\rm SFR}_{\rm 10}\geq0.3\,\msunyr$ are selected to simulate emission lines (see Table~\ref{tab:sample} for more details). The typical galactic stellar masses we probe with the SFR criterion range between $7 \la \log_{10}({\rm M_{\rm *,i}}/{\rm M_{\odot}}) \la 9$ at $z>7$, which are comparable to those from the early JWST results \citep[e.g.,][]{Santini2023}. More massive galaxies with stellar masses up to  $4\times10^{10}\ {\rm M_{\odot}}$ appear in later epochs ($z=4.64$), providing the opportunity to learn more about the physical nature of Ly$\alpha$ emitters, such as those observed with MUSE \citep[e.g.][]{Bacon2017}. 

Our choice of SFR cut is motivated by the recent analysis from JWST observations \citep[][]{Leethochawalit2023,Fujimoto2023,Shapley2023} that are subject to a flux limit. This of course ignores the fact that UV luminosity and SFR do not necessarily map one-to-one (primarily because of dust attenuation); however, we cannot determine the UV luminosity until after post-processing (which is the computationally limiting step). Our cut excludes a large fraction of galaxies with ${\rm M_{*,i}}\la 10^{8}\ {\rm M_{\odot}}$. For example, our galaxy samples are 90\% complete in stellar mass down to $\log_{10}({\rm M_{*,i}}/[{\rm M_{\odot}}])=8.0$, $8.5$, $8.9$, and $8.8$ at $z=10$, $8$, $6$, and $4.64$, respectively, and thus are biased (by construction) towards star-forming galaxies. This should be kept in mind in the following sections when we compare \sphinx{} data to observations.

\subsubsection{Stellar mass - Halo mass relation}
\begin{figure}
\centering
\includegraphics[width=0.45\textwidth]{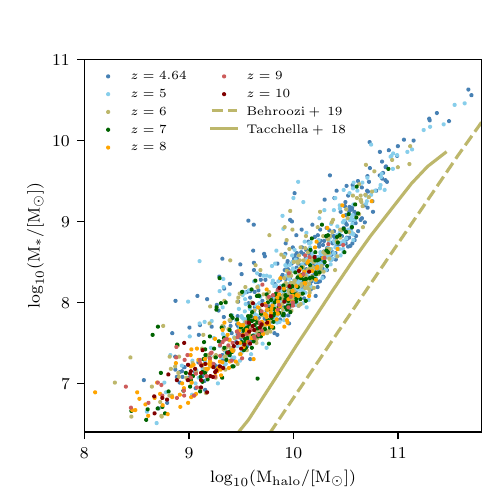}
  \caption{Stellar mass-halo mass relation for galaxies in the \sphinx{} data release compared to the $z=6$ model of \protect\cite{Tacchella2018} and $z=6$ estimates from abundance matching \protect\cite{Behroozi2019}.}
  \label{fig:m_star_m_halo}
\end{figure}

Although neither quantity is actually observable, the stellar mass-halo mass relation provides important information on the galaxy-halo connection at all redshifts and constraints from abundance matching are often used to tune subgrid feedback models in simulations. This relation has been presented for \sphinx{} data for all galaxies in \cite{Rosdahl2018,Rosdahl2022} and is available for star-forming galaxies as part of the \sphinx{} data release. We show our results in Figure~\ref{fig:m_star_m_halo}. For comparison we also plot the stellar mass-halo mass relation from the model of \cite{Tacchella2018} as well as from abundance matching \citep{Behroozi2019}, both of which are estimated at $z=6$. 

We note a few important points regarding high-redshift stellar mass. Despite its ubiquity as a constraint on feedback, stellar mass at these redshifts is not an observed quantity; hence, constraints from observational data are model-dependent. This is particularly true at high-redshift where the JWST photometry primarily probes the rest-frame UV and shorter-wavelength optical and HST data probes only rest-frame UV. These regions of the spectrum are dominated by young stellar populations, especially if the galaxies exhibit bursty star formation histories and thus the prior on the star formation history is a key parameter in determining the final stellar mass (in addition to stellar IMF and various other quantities). This is reflected by the fact that spatially resolved stellar mass estimates at high-redshift often result in stellar masses that are $\sim0.5-1$~dex higher than the galaxy-integrated photometry \citep{Clara2023}. This offset is consistent with the difference between the \sphinx{} prediction and that from abundance matching. The \sphinx{} data release can perhaps play a role in improving the priors used in SED fitting codes (see also \citealt{Narayanan2023}). Finally the models of \cite{Tacchella2018,Behroozi2019}, which have been used as benchmarks for theoretical models of high-redshift galaxy formation, do not reproduce the cumulative number counts of high-redshift galaxies from JWST (see e.g. \citealt{Leung2023} and Section~\ref{sec:rt_comp_no_ism}), neither do the simulations that have been tuned to reproduce them \citep[e.g.][]{Kannan2022}. 

\subsubsection{Star formation rates} 
\label{sec:sfr_vs_sfr}

\begin{figure}
\centering
\includegraphics[width=0.45\textwidth,trim={0 1.5cm 0 0},clip]{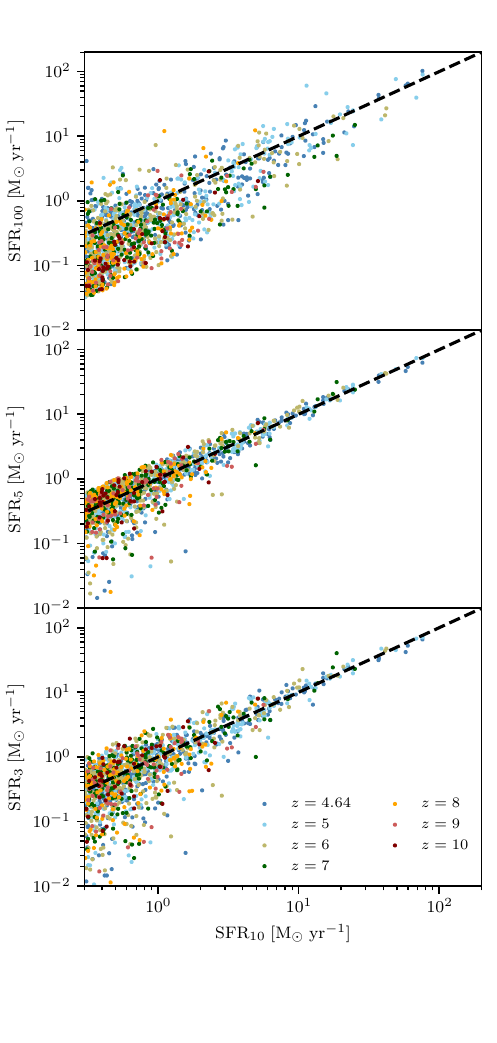}
  \caption{Star formation rates averaged over 100~Myr ($\rm SFR_{100}$, top), 5~Myr ($\rm SFR_{5}$, middle) and 3~Myr ($\rm SFR_{3}$, bottom) as a function of 10~Myr-averaged SFR ($\rm SFR_{10}$). Individual points represent \sphinx{} galaxies at different redshifts, selected to have $\rm SFR_{\rm 10}\geq0.3\,M_{\odot}\ yr^{-1}$. The black diagonal lines show the one-to-one relations.}
  \label{fig:sfr_averages}
\end{figure}
Observationally, different tracers exist for determining the SFR of a galaxy such as UV and IR luminosity as well as various emission lines across the wavelength spectrum \citep[e.g.][]{Kennicutt1998, Kennicutt2012,delooze2014}. Different SFR indicators probe star formation on various timescales. For instance, the ${\rm H}\alpha$ recombination line results from gas ionised by young stars, and hence gives a hint of the current SFR on $\lesssim10$~Myr time scales. In contrast, FUV continuum emission traces a longer-time scale SFR, up to 100~Myr. Therefore, the SFR from a galaxy spectrum may differ depending on the SFR indicator that is used. In order to illustrate the consequences of the star formation history variability, we provide in \spdrone\ SFRs measured over multiple time scales.

Figure~\ref{fig:sfr_averages} shows SFR averaged over 100, 5, and 3~Myr (respectively $\rm SFR_{100}$, $\rm SFR_{5}$, and $\rm SFR_{3}$) as a function of 10~Myr-averaged SFR ($\rm SFR_{10}$). All the galaxies in the data release are selected to have $\rm SFR_{\rm 10}\geq0.3\ M_{\odot}\ yr^{-1}$, but on different time scales, the SFR can drop well below this value. For example, on average, $\rm SFR_{100}$ tends to fall well below $\rm SFR_{10}$, indicating that the 10~Myr-averaged SFR is more representative of a burst rather than the continuous star formation history. In contrast, $\rm SFR_{5}$ and $\rm SFR_{3}$ are more evenly scattered around the one-to-one relation. However, there are a few galaxies with very low $\rm SFR_{5}$ and $\rm SFR_{3}$, which is indicative of a short-timescale quenching event. 

\subsubsection{Star formation main sequence} \label{sec:sfr_vs_mstar}
\begin{figure}
\centering
\includegraphics[width=0.45\textwidth]{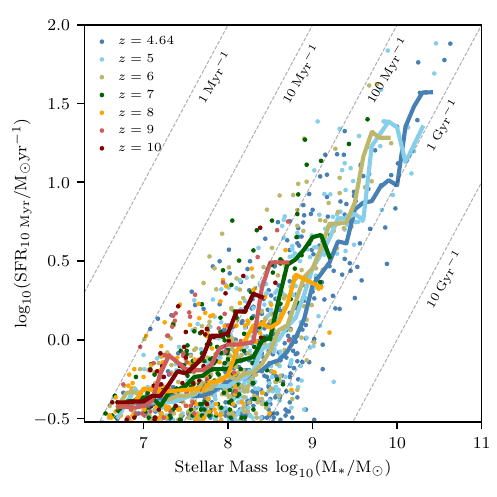}
\includegraphics[width=0.45\textwidth]{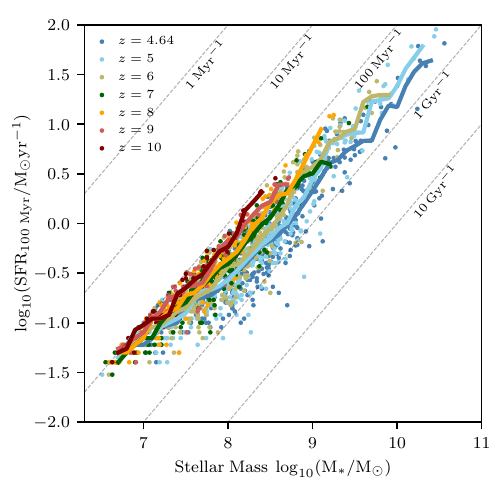}
  \caption{Star formation main sequence in \sphinx{} for different redshifts for 10~Myr and 100~Myr averaged SFRs (upper and lower panel, respectively). The points represent individual galaxies and the solid lines represent median star formation rates. The alignment of points at low SFR is due to the formation of integer numbers of star particles.}
  \label{fig:sfms}
\end{figure}

There is a well established relationship between the stellar mass of a galaxy and its SFR \citep[e.g.][]{Brinchmann2004,Salim2007} that extends to the epoch of reionization \citep[e.g.][]{Popesso2023}. \sphinx{} galaxies similarly exhibit such a relationship. In Figure~\ref{fig:sfms} we show both the 10~Myr-averaged (top) and 100~Myr-averaged (bottom) SFRs as a function of stellar mass. The SFRs in the \sphinx{} data release reach as high as 100~${\rm M_{\odot}\ yr^{-1}}$ and span more than three~dex in SFR on longer timescales. On 100~Myr timescales there is a tight correlation such that specific SFRs (sSFRs) typically fall between $10^{-8}-10^{-9}$~yr$^{-1}$ while on 10~Myr timescales sSFRs can go as high as $10^{-7}$~yr$^{-1}$. Such high values are consistent with some of the most highly star-forming extreme emission line galaxies at lower redshift \citep[e.g.][]{Amorin2017} as well as tentative estimates from early JWST observations \citep[e.g.][]{Shapley2023,Curti2023}. The high sSFRs are clearly not sustained over long timescales due to the fact that the 100~Myr-averaged SFRs are in general lower than the 10~Myr-averaged values for our most highly star-forming galaxies.

We note that the flattening in the main sequence at the low-mass end of the 10~Myr-averaged figure is entirely due to the threshold SFR used to create the \sphinx{} public data set. If all galaxies were included, our relations would trend towards lower SFR, albeit the relation would still exhibit significantly more scatter than the 100~Myr-averaged main-sequence. This is an important point to consider, and a recurring theme throughout this work that higher redshift galaxies in \spdrone\ tend to probe galaxies further above the main sequence. In many ways this is reflective of the behaviour of a flux-limited survey where the fainter galaxies at higher-redshifts must fall consistently higher above the main-sequence to be observed.

\subsubsection{Star formation histories}
\label{sec:sfh}
\begin{figure}
\centering
\includegraphics[width=0.45\textwidth]{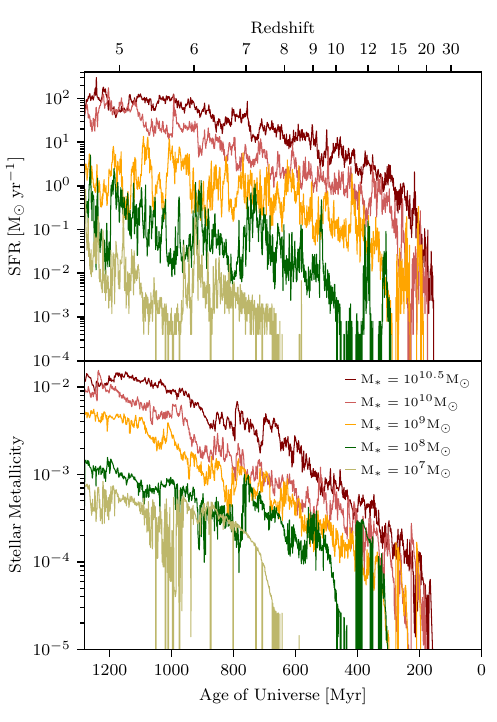}
  \caption{Star formation histories (top) and stellar metallicity histories (bottom) of five \sphinx{} galaxies, with stellar masses at $z=4.64$ as indicated in the legend.}
  \label{fig:sfh}
\end{figure}

The \sphinx{} data release contains the star formation histories (SFHs) of all sampled halos, with a time resolution of 1~Myr. We show in Figure~\ref{fig:sfh} the SFHs of five example galaxies in the $z=4.64$ snapshot, selected to span the range of galaxy masses in our sample, as indicated in the plot legend. As is typical for all \sphinx{} galaxies, the SFHs shown are very stochastic and bursty, increasingly so for decreasing galaxy stellar masses. The plot also demonstrates that the lower-mass galaxies have spent most of their lives with star formation rates below the selection limit of the data release sample, i.e. with ${\rm SFR\geq0.3\ M_{\odot}\ yr^{-1}}$, and therefore would approximately phase in and out of being observable with the JWST. One important property of these SFHs is that they represent star formation in all of the progenitor haloes, rather than the main branch along the merger tree. This is important for SED modelling as the total star formation in all progenitors is the quantity constrained by the integrated SED.

\begin{figure}
\centering
\includegraphics[width=0.45\textwidth]{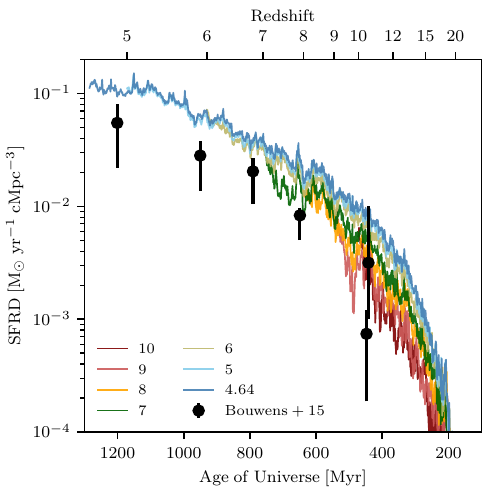}
\includegraphics[width=0.45\textwidth]{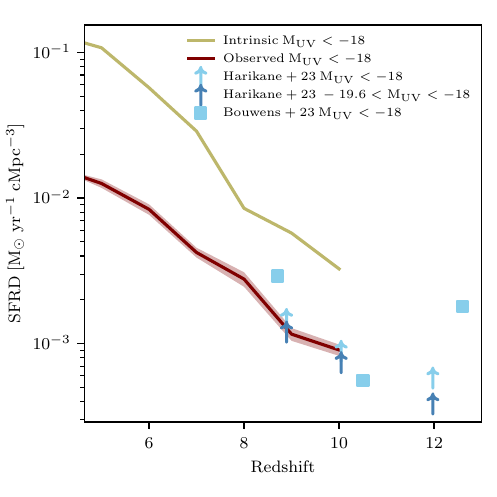}
  \caption{(Top) SFRD of the \sphinx{} volume, using the galaxies selected for the data release, i.e. traced back in time for all galaxies with ${\rm SFR_{10}\geq0.3 \ \msunyr}$ from \sphinx{} snapshots at the redshifts indicated in the legend. For comparison we show the dust-corrected estimates for the SFRD from \citet{Bouwens2015}. (Bottom) SFRD of the \sphinx{} volume measured from the UV magnitude of each \sphinx{} galaxy at each redshift using either the intrinsic ${\rm M_{UV}}$ (yellow) or dust obscured ${\rm M_{UV}}$ (red) down to a limiting magnitude of $-18$. For comparison we show observational constraints from \protect\cite{Harikane2023,Bouwens2023} that have been integrated to the same limiting magnitude. We have converted their reported UV luminosity density into an SFRD using our Equation~\ref{eqn:muv_sfr}.}
  \label{fig:sfh_all}
\end{figure}

We can also extract the SFR density of the early universe from the \sphinx{} data. This is shown in the top panel of Figure~\ref{fig:sfh_all}, where we plot the SFR density traced back in time for all sampled galaxies (i.e. with ${\rm SFR\geq0.3 \ \msunyr}$) at the given redshifts. Note how the total SFRD for galaxies in our $z=4.64$ catalogue always tracks above those in the higher redshift catalogues. This is due to the fact that the low mass progenitors that do not reach our SFR threshold merge into larger haloes by $z=4.64$. Thus their star formation is included in the history of the massive galaxies in the lower-redshift snapshots but not at high redshift.

This intrinsic SFRD is, however, not directly comparable to observations, which instead measure a UV luminosity density ($\rho_{\rm SFR}$) above some magnitude threshold that is then converted into an SFRD. To demonstrate this comparison, we first derive an empirical relation between absolute magnitude at 1500~\AA, M$_{\rm UV,1500}$, and 10~Myr-averaged SFR from our data set such that
\begin{equation}
\label{eqn:muv_sfr}
    \log_{10}\left(\frac{\rm SFR_{10}}{\rm M_{\odot}\ yr^{-1}}\right) = -0.35\,{\rm M_{UV,1500}} - 6.49.
\end{equation}
We then select all galaxies that have an intrinsic M$_{\rm UV}\leq-18$ and calculate their total SFRD at each snapshot. This is shown as the yellow line in the bottom panel of Figure~\ref{fig:sfh_all}. We then use the observed M$_{\rm UV}$ values along each of our ten sight lines (i.e. those uncorrected for dust) and convert to an SFRD using Equation~\ref{eqn:muv_sfr}. The red line and shaded region in the bottom panel of Figure~\ref{fig:sfh_all} represent the SFRD that an observer would derive for our sample at each snapshot for galaxies with observed M$_{\rm UV}\leq-18$. At $z=4.64$, the observed and true SFRDs differ by an order of magnitude due to obscuration by dust. For comparison, we have converted $\rho_{\rm UV}$ for spectroscopically confirmed high-redshift JWST galaxies with M$_{\rm UV}\leq-18$ from \citet{Harikane2023} to an SFRD using our Equation~\ref{eqn:muv_sfr} (for consistency). Because these measurements do not correct for dust and are only spectroscopically confirmed galaxies, they constitute a firm lower-limit on the SFRD. These are shown at the light blue arrows in the bottom panel of Figure~\ref{fig:sfh_all} and they are surprisingly consistent with our mock observations. We emphasize that this agreement is to some extent serendipitous because \sphinx{} does not probe galaxies as bright as seen in \citet{Harikane2023} due to its limited simulation box size. For example, the brightest dust-obscured $z=10$ galaxy in \sphinx{} has a UV magnitude of -19.6 while \citet{Harikane2023} probe galaxies brighter than $-21$ at this redshift. Selecting only fainter galaxies from the \citet{Harikane2023} sample (i.e. those with ${\rm M_{UV}>-19.6}$) inevitably decreases the SFRD as shown by the dark blue arrows. We have performed the same experiment using data from \cite{Bouwens2023} which has a similar magnitude limit and the results differ compared to the \cite{Harikane2023} data. This may simply be due to cosmic variance or different estimates of survey volume. Nevertheless, the key feature is that even at such high redshifts, dust plays an important role in obscuring star formation and the \sphinx{} data set provides a means to directly compare SFRD estimates between simulations and observations.

\subsubsection{Stellar ages}
\begin{figure}
\centering
\includegraphics[width=0.45\textwidth]{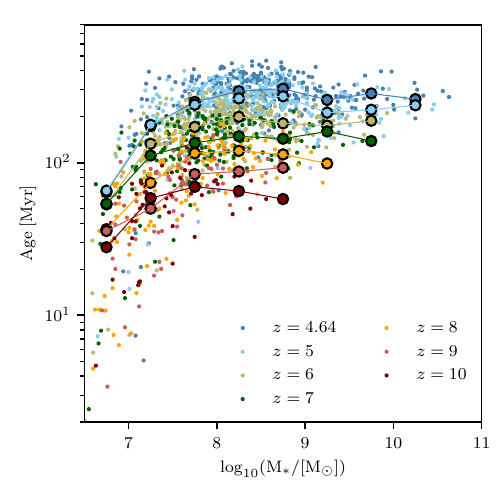}
  \caption{Mass-weighted stellar age in \sphinx{} for different redshifts as a function of galaxy stellar mass. The solid lines and large data points indicate the mean values while smaller data points show individual galaxies.}
  \label{fig:stellar_age}
\end{figure}

The stellar age of a galaxy is an important parameter that encapsulates its mass growth history. Mass-weighted stellar ages are provided in \spdrone\ and can likewise be recomputed from the SFHs. Figure~\ref{fig:stellar_age} shows the mass-weighted stellar ages as a function of stellar mass. Interestingly, at ${\rm M_*\gtrsim 10^{7.5}\ M_{\odot}}$ the trend is relatively flat at a fixed redshift such that galaxies, independent of their stellar mass, exhibit similar stellar ages. However, as redshift increases, ages of galaxies decrease. At lower stellar masses, the galaxies tend to exhibit younger stellar ages but indeed this is due to the SFR threshold in the database. At such low stellar masses, a 10~Myr-averaged SFR of $0.3\ {\rm M_{\odot}\ yr^{-1}}$ represents a stellar mass of $3\times10^6\ {\rm M_{\odot}}$, which is a significant fraction of the total stellar mass of objects with ${\rm M_*< 10^{7.5}\ M_{\odot}}$

\begin{figure}
\centering
\includegraphics[width=0.45\textwidth]{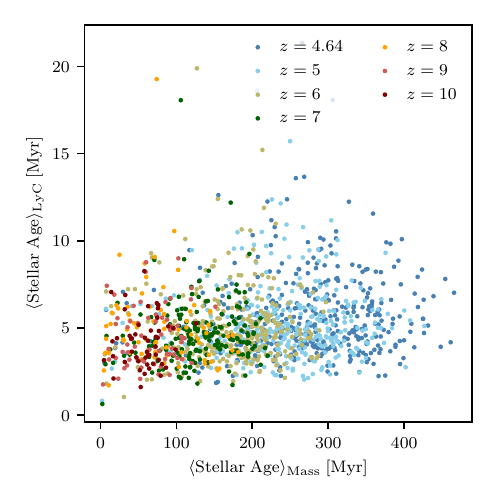}
  \caption{LyC-weighted stellar age versus mass-weighted stellar age for \sphinx{} galaxies in the data release.}
  \label{fig:age_mass_lyc}
\end{figure}

Stellar age need not be defined as a mass-weighted quantity. At high-redshift, JWST is primarily sensitive to the rest-frame UV and optical parts of a spectrum. Furthermore, the emission from nebular regions cares only about the stars that emit LyC photons. For this reason, we also provide stellar ages weighted by LyC emission. A comparison between the mass-weighted and LyC-weighted ages is shown in Figure~\ref{fig:age_mass_lyc}. While the mass-weighted ages reach values of hundreds of Myr and represent a significant fraction of the Hubble time at each particular redshift, the LyC-weighted ages are only sensitive to the young stellar populations and hence probe the recent star formation in the past $\sim10$~Myr. 

\subsubsection{Stellar metallicity histories}
In addition to the SFHs, we also provide stellar metallicity histories for each galaxy in the \sphinx{} data release. These are computed as the mass-weighted metallicity of all star particles that form in the same 1~Myr bins as were used to compute the SFHs. The bottom panel of Figure~\ref{fig:sfh} shows the stellar metallicity evolution for the same five galaxies as in the top panel. Note that the stellar metallicity functions are not monotonic. Pristine inflows can dilute the gas metallicity of the galaxies which is then imprinted on the stars. Similarly, the metallicity evolution contains the star formation events in all  progenitors with different metallicity evolution and SFHs. Similar to the SFHs, the metallicity evolution is a key component of the SED modelling, and thus the \sphinx{} data can be used as a prior when analyzing real observations.

\subsubsection{Mass-metallicity relation}
The stellar mass/gas-phase metallicity relation (MZR) is one of the primary galaxy scaling relations that encapsulates information on the baryon cycle. Hence, constraining this relation is one of the primary goals of high-redshift JWST observations \citep[e.g.][]{Nakajima2023,Curti2023}. The \sphinx{} data release contains all of the information needed to compare simulated metallicity to that which can be probed with observations.

\begin{figure}
\centering
\includegraphics[width=0.45\textwidth,trim={0.0 1.3cm 0.0 0.0},clip]{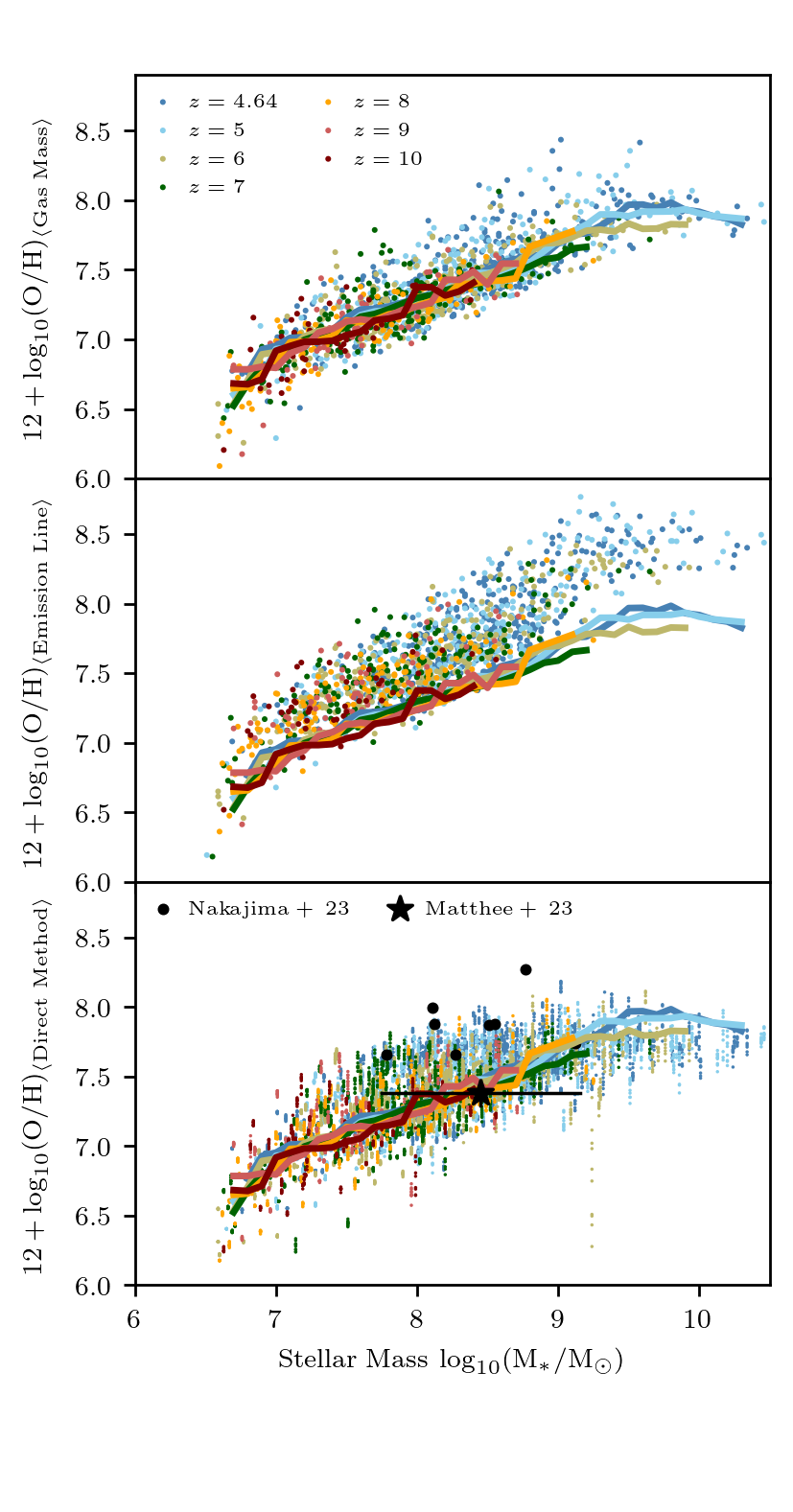}
  \caption{Stellar mass/gas-phase metallicity relation for SPHINX$^{20}$ galaxies computed in three different ways. The top, middle, and bottom panels show the gas mass-weighted metallicity, the intrinsic [O~III]~$\lambda$5007-, [O~II]~$\lambda\lambda$3727-, and H$\beta$-weighted metallicity, and the metallicity measured using the direct method along each viewing angle, respectively. The different colours represent galaxies at different redshifts and the solid lines on all panels represent running median fits to the mass-weighted metallicity shown in the top panel. In the bottom panel, we show seven high-redshift galaxies with [O~III]~$\lambda$4363 detections from \protect\cite{Nakajima2023} as black circles where we have recomputed metallicities to be consistent with the atomic data used in \sphinx{}. We also show the stacked data with [O~III]~$\lambda$4363 detections from \protect\cite{Matthee2023} based on JWST NIRCam grism data.}
  \label{fig:mzr}
\end{figure}

In the top panel of Figure~\ref{fig:mzr} we show the MZR where metallicity is weighted by gas mass within each halo. There is little evolution with redshift among our star-forming sample. This lack of evolution is partially driven by the fact that the SFR cut means that the higher redshift galaxies tend to fall further above the star formation main-sequence. However we emphasize that while the mass-weighted metallicity is the value predicted by most cosmological simulations, this is not a quantity that is accessible to observers. Rather, observations are only sensitive to metallicity in the bright, line-emitting H~{\small II} regions. In the middle panel we show the metallicity weighted by the [O~III]~$\lambda$5007, [O~II]~$\lambda\lambda$3727, and H$\beta$ luminosities of the gas cells, which better represents the metallicity of H~{\small II} regions. These values are significantly higher than the mass-weighted metallicity of the galaxy demonstrating the early self-enrichment of galaxies. 

Because the data release contains all of the necessary emission lines to apply the direct method, which is widely considered the gold-standard metallicity estimator, in the bottom panel of Figure~\ref{fig:mzr} we show the MZR where metallicity is computed along each sight line. Specifically, we use the reddening corrected values of [O~III]~$\lambda$5007, [O~III]~$\lambda$4363, [O~II]~$\lambda\lambda$3727, and H$\beta$ and use the \cite{Pilyugin2009} conversion between t$_{\rm O3}$ and t$_{\rm O2}$\footnote{t$_{\rm O3}$ and t$_{\rm O2}$ correspond to the electron temperature in the high and low ionization zones, respectively.}. We reproduce the result from \cite{Cameron2022} where H~{\small II} region metallicity is underpredicted by the direct method due to temperature fluctuations between H~{\small II} regions across the simulated galaxies. The \sphinx{} data release contains auroral lines for many elements and ionization states so that the direct method metallicities can be compared between \sphinx{} and observations for numerous JWST programs (e.g. ID 2953 - CECILIA - PIs Strom, Rudie, ID 1914 - AURORA - PIs Shapley, Sanders, ID 1879 - PI Curti).

\subsubsection{ISM conditions}
\begin{figure}
\includegraphics[width=0.45\textwidth]{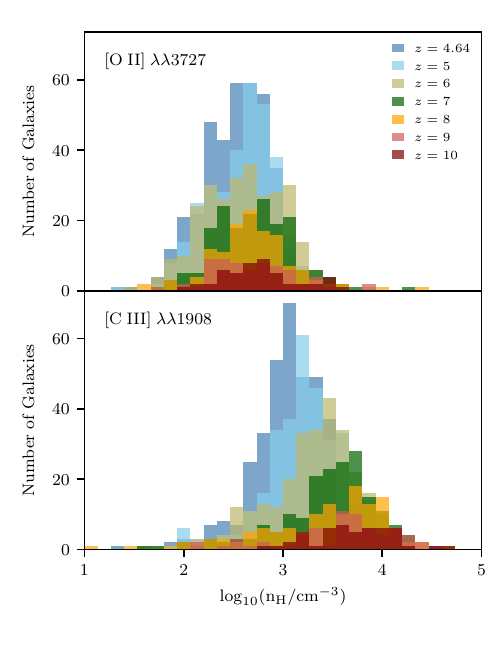}
\caption{Histograms of ISM gas densities weighted by intrinsic [O~{\small II}]~$\lambda\lambda$3727 (top) or [C~{\small III}]~$\lambda\lambda$1908 (bottom) emission as a function of redshift.}
\label{fig:ism_density}
\end{figure}

NIRSpec access to emission lines at high-redshift means that certain ISM properties such as gas density, metallicity, ionization parameter, temperature, etc. can be inferred. As we showed above, this is paramount for constraining the mass-metallicity relation; although, the emission lines provide a biased view of the metallicity of the galaxy. The \sphinx{} data release also contains various other ISM properties including gas density. In Figure~\ref{fig:ism_density} we show histograms of ISM density in each redshift bin weighted by intrinsic [O~{\small II}]~$\lambda\lambda$3727 (top) or [C~{\small III}]~$\lambda\lambda$1908 (bottom) emission. We highlight two interesting features. First, [C~{\small III}]~$\lambda\lambda$1908 emission is generally probing higher gas densities than [O~{\small II}]~$\lambda\lambda$3727 in our galaxies. Second, the [O~{\small II}]~$\lambda\lambda$3727 gas density has little redshift evolution while the [C~{\small III}]~$\lambda\lambda$1908 seems to increase with increasing redshift. This is due to the fact that the different ionization potentials of O$^+$ and C$^{++}$ mean that the emission lines originate from different regions of the nebula. These points are both crucial for interpreting the evolution of ISM properties with multi-line tracers and hence the \sphinx{} data release can be a valuable tool in this regard as standard photoionization models cannot capture this behaviour (see discussion in \citealt{Katz2022-CII-OIII}).

\subsubsection{LyC escape fractions}
\begin{figure}
\includegraphics[width=0.45\textwidth]{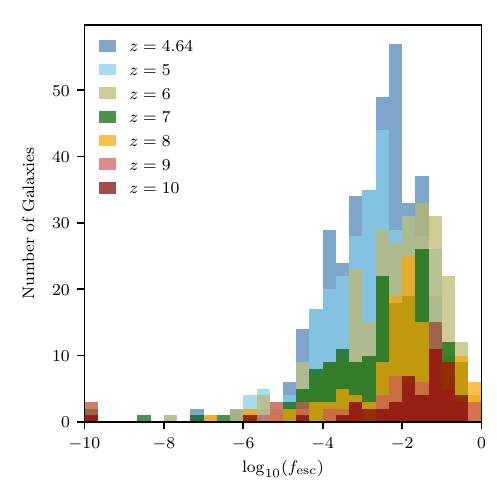}
\caption{Angle-averaged LyC escape fractions for \sphinx{} galaxies as a function of redshift.}
\label{fig:fesc}
\end{figure}

Constraining the sources that reionized the Universe is one of the primary science goals of JWST. Because of its spatial resolution and multiphase ISM, \sphinx{} is one of the few full-box radiation hydrodynamics simulations that can predict escape fractions\footnote{These predictions are valid on the scale of the spatial resolution of the simulation (i.e. $\sim10$~pc).}. LyC escape fractions from \sphinx{} have been discussed at-length in previous works \citep{Rosdahl2018,Rosdahl2022,Choustikov2023} and hence we refrain from further discussion here. Methods for computing LyC escape fractions can be found in the aforementioned references. The data set contains both angle-averaged $f_{\rm esc}$ (which includes all photons with $E>13.6\ {\rm eV}$) as well as those along the fiducial ten sight lines (for photons specifically at 900~\AA). In Figure~\ref{fig:fesc} we show histograms of the angle-averaged LyC escape fraction (i.e. the value that is important for reionization) for each redshift. The values range from nearly 100\% to zero. We highlight the tendency for the typical $f_{\rm esc}$ of star-forming galaxies to decrease with decreasing redshift, as is discussed at length in \cite{Rosdahl2022}.

\subsubsection{Ionizing photon production efficiency}
\begin{figure}
\includegraphics[width=0.45\textwidth]{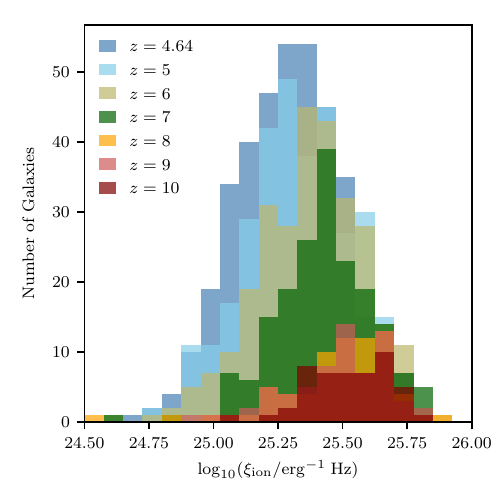}
\caption{Histograms of intrinsic $\xi_{\rm ion}$ for the different redshift snapshots available in the \sphinx{} data release.}
\label{fig:xi_ion}
\end{figure}

Similarly important for reionization as $f_{\rm esc}$ is the ionizing photon production efficiency, $\xi_{\rm ion}\equiv Q_{\rm LyC}/L_{\rm UV}$, where $Q_{\rm LyC}$ is the production rate of LyC photons and $L_{\rm UV}$ is the monochromatic luminosity at 1500~\AA. Early JWST observations have attempted to constrain this value for star-forming galaxies during the epoch of reionization \citep[e.g.][]{Simmonds2023}. While the ionizing photon production efficiency of each star particle is given by its age and metallicity from our chosen \bpass{} SED, the complex SFHs of our galaxies mean that distributions of $\xi_{\rm ion}$ and their evolution are non-trivial. Furthermore, comparing the observed $\xi_{\rm ion}$ with the intrinsic $\xi_{\rm ion}$ is key for interpreting how well observations constrain this quantity. We refrain from that exercise here and show in Figure~\ref{fig:xi_ion} the intrinsic distribution of $\xi_{\rm ion}$ for galaxies in the \sphinx{} data release. The typical values increase with increasing redshift as we sample lower metallicity galaxies further above the main-sequence at higher redshift. Due to our adopted SED, our values do not strongly deviate from those typically assumed in reionization models \citep[e.g.][]{Robertson2015}. 

\subsection{Photometry and Imaging}
In the previous section we primarily focused on the intrinsic galaxy properties that are distributed as part of \spdrone\ as well as how they relate to certain observable quantities. In this section, we continue our demonstration of the database by highlighting data that is relevant to compare with imaging and aperture photometry.

\subsubsection{Photometry}
\label{sec:phot}
\begin{figure}
\includegraphics[width=0.45\textwidth,trim={0 0.5cm 0 0.2cm},clip]{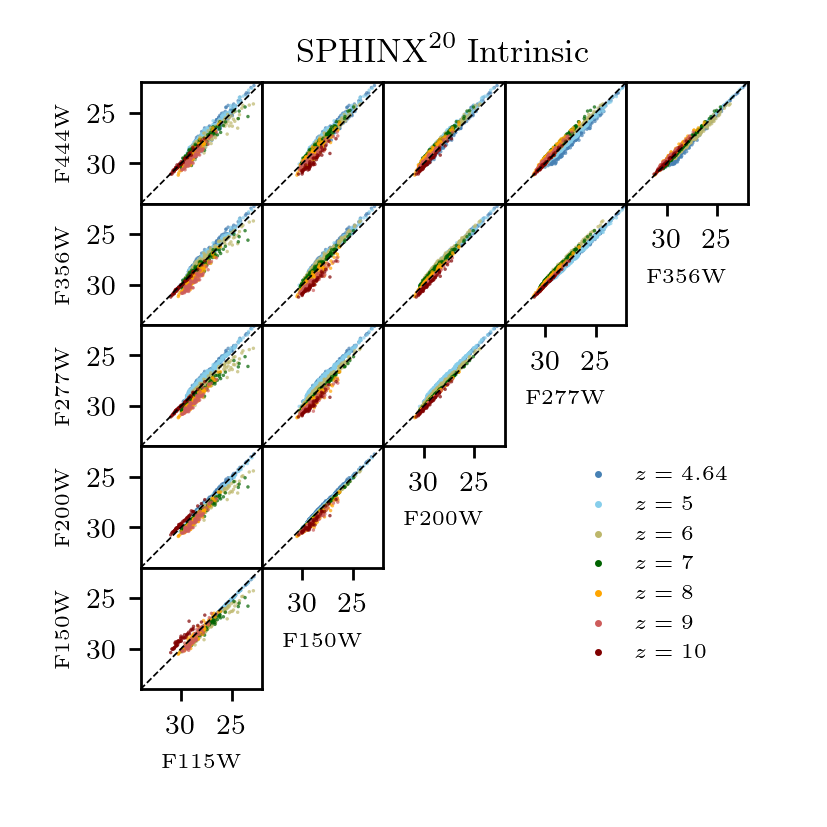}
\includegraphics[width=0.45\textwidth,trim={0 0.5cm 0 0.2cm},clip]{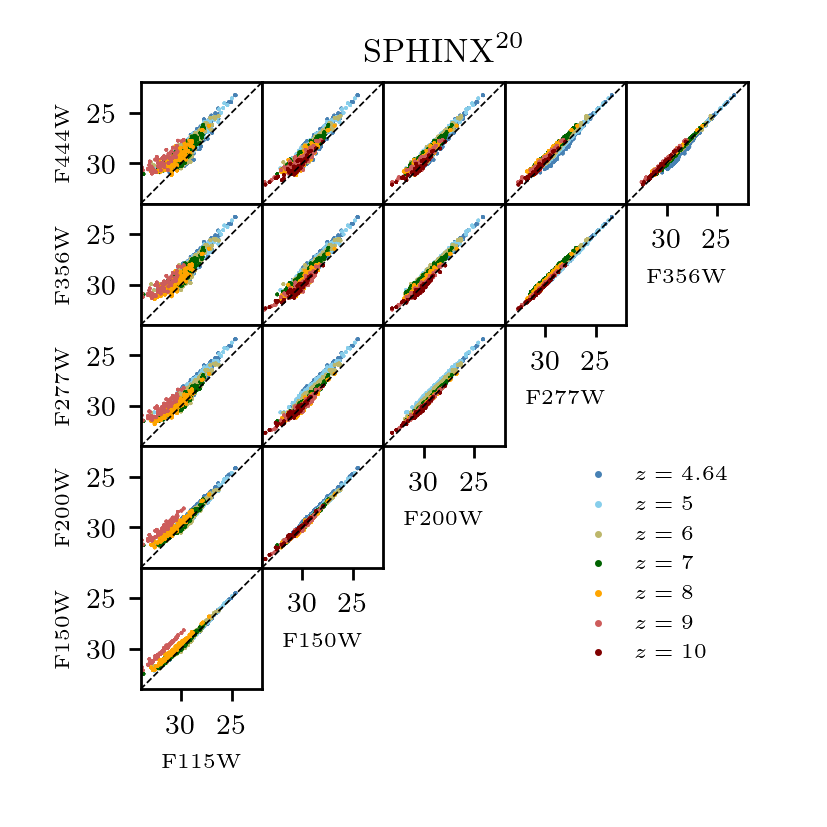}
\includegraphics[width=0.45\textwidth,trim={0 0.5cm 0 0.2cm},clip]{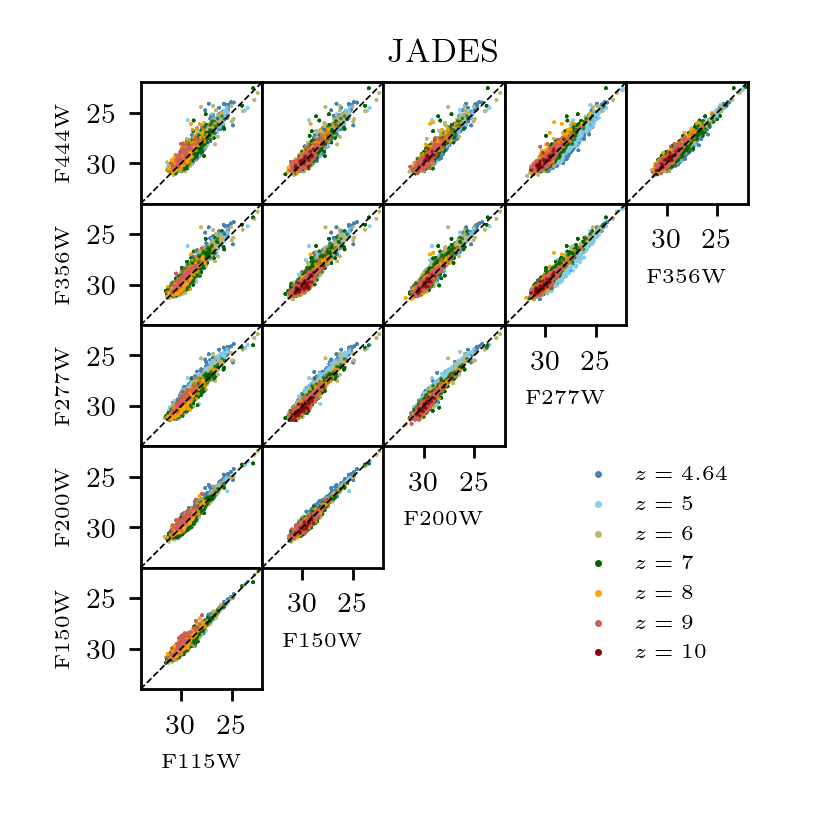}
\caption{Distribution of NIRCam wide filter magnitudes for \sphinx{} galaxies using their intrinsic magnitudes (Top), dust-attenuated magnitudes (Middle), and for galaxies observed as part of the JADES GTO program (Bottom). JADES galaxies are selected based on photometric (or spectroscopic where available) redshift. The dashed black lines represent the one-to-one relation.}
\label{fig:colour-hist}
\end{figure}

The \sphinx{} data can be used to make mock photometric catalogs in any filter that covers the rest-frame UV and optical\footnote{All photometry released in \spdrone\ is computed from the spectra with {\small sedpy} (\url{https://github.com/bd-j/sedpy}).}. We demonstrate this in Figure~\ref{fig:colour-hist} where the top and middle panels show the filter magnitudes of \sphinx{} galaxies for all NIRCam wide filters plotted against each other for intrinsic and dust attenuated emission, respectively. For comparison, the bottom panel shows the same relations for JADES galaxies from the publicly available catalog \citep{Bunker2023b,Eisenstein2023,Hainline2023,Rieke2023} selected by photometric redshift (or spectroscopic when available) and binned by similar redshifts as the \sphinx{} data. We only consider galaxies with a signal-to-noise ratio $>3$ in each of the filter combinations.

The dust-attenuated \sphinx{} galaxies cover a very similar magnitude distribution as what is observed in JADES, making it an ideal comparison sample. Nevertheless, there are some differences. First, JADES contains some brighter galaxies that are not present in the \sphinx{} data release due to the finite volume of the simulation. Similarly, although we implement an SFR threshold, the \sphinx{} data release contains galaxies below the magnitude limit of JADES. The most notable difference is that the observed galaxies are bluer than those in \sphinx{}. For example, comparing F115W versus F444W, many JADES galaxies fall below the one-to-one relation indicating they are very blue.  

By comparing the top and middle panels of Figure~\ref{fig:colour-hist}, we can gain insight into the origin of this discrepancy. Without dust, many of the \sphinx{} galaxies fall below the one-to-one relation and are thus more consistent with JADES. However, without dust the scatter in the \sphinx{} data is likely too small compared to observations. One way to solve this it to adopt either a dust attenuation law that is less steep with wavelength compared to the SMC. For example the \cite{Reddy2015} and \cite{Calzetti2000} curves exhibit such behaviour. One could also adopt a dust-to-gas mass ratio that falls off as a power-law with metallicity rather than linearly, which is perhaps more supported by observations \citep[e.g.][]{RR2014}. A more top-heavy IMF can also make the galaxies appear bluer \citep[e.g.][]{Stanway2023}, and simultaneously solves various other issues related to the number of observed bright galaxies \citep[e.g.][]{Yung2023} and ALMA line emission \citep[e.g.][]{Katz2022-CII-OIII}. Finally, a reduction in total stellar mass (e.g. stronger feedback) would preferentially reduce the flux at longer wavelengths.

From the observational perspective, the galaxies with the highest sSFRs are likely to be most easily detected so there might be a small bias towards seeing bluer galaxies in observations, or some of the galaxies may be low-redshift interlopers. At $z=9$, the Ly$\alpha$ break falls in the F115W filter, so one might expect that the pink points in Figure~\ref{fig:colour-hist}, representing the $z=9$ galaxies, would be slightly offset to the left, e.g. from the $z=8$ sample, if there is no strong spectral evolution. This is not true for the JADES data where the $z=9$ photometrically selected sample overlaps with the other, lower-redshift galaxies. Hence the observed sample must be increasingly blue with redshift as reported in \cite{Topping2023}. 

Finally, one should note the importance of emission lines. At $z=9$, [O~{\small III}]~$\lambda$5007 and H$\beta$ fall in the F444W filter. If the emission lines in \sphinx{} are too strong, our galaxies may appear too red. However, simulations generally struggle to produce the observed tail of high equivalent widths of [O~{\small III}]~$\lambda$5007$+$H$\beta$ \citep[e.g.][]{Wilkins2023b} so we believe this is unlikely to cause the discrepancy. Furthermore, \sphinx{} galaxies appear too red compared to JADES galaxies in other filters besides F444W and F115W. If many of the JADES galaxies are Ly$\alpha$ emitters, this could similarly boost the F115W flux but not solve the issue of \sphinx{} galaxies being too red in F150W, unless other UV emission lines are similarly as bright. However such bright UV lines are simply not observed in either \sphinx{} or spectroscopically confirmed JADES galaxies \citep{Curtis-Lake2022}. Further analysis that is beyond the scope of this paper will be required to elucidate the origin of the differences between \sphinx{} and JADES.

\subsubsection{Colour-colour selection}
Colour selections designed to capture the Lyman break represent one of the primary techniques for photometrically selecting samples of high-redshift galaxies \citep[e.g.][]{Steidel1996,Giavalisco2002}. Understanding how well certain colour selections work in terms of false positive and false negative rates are key for accurate determinations of luminosity functions and for follow-up spectroscopic observations. The \sphinx{} public data release contains magnitudes for all JWST filters along each sight line and filter magnitudes for any other telescope can be trivially computed from the full spectra provided (see below). 

In Figure~\ref{fig:colour-colour} we show an example colour-colour diagram and plot F115W$-$F150W versus F150W$-$F277W for each galaxy at each redshift. The black line demarcates the colour-colour selection from \cite{Harikane2023} for $z=9$ galaxies. It is clear that the vast majority of \sphinx{} $z=9$ galaxies are selected by this criteria; however, there is some contamination from $z=8$ systems and a handful of $z=9$ galaxies fall outside the selection boundary. For comparison, we also show the locations of JADES galaxies in the plane and highlight the ones with photometric or spectroscopic redshifts $>9$. The realistic spectra in the \sphinx{} public data release can be used to optimize these criteria in combination with observations of potential lower-redshift contaminants.

\begin{figure}
\includegraphics[width=0.45\textwidth]{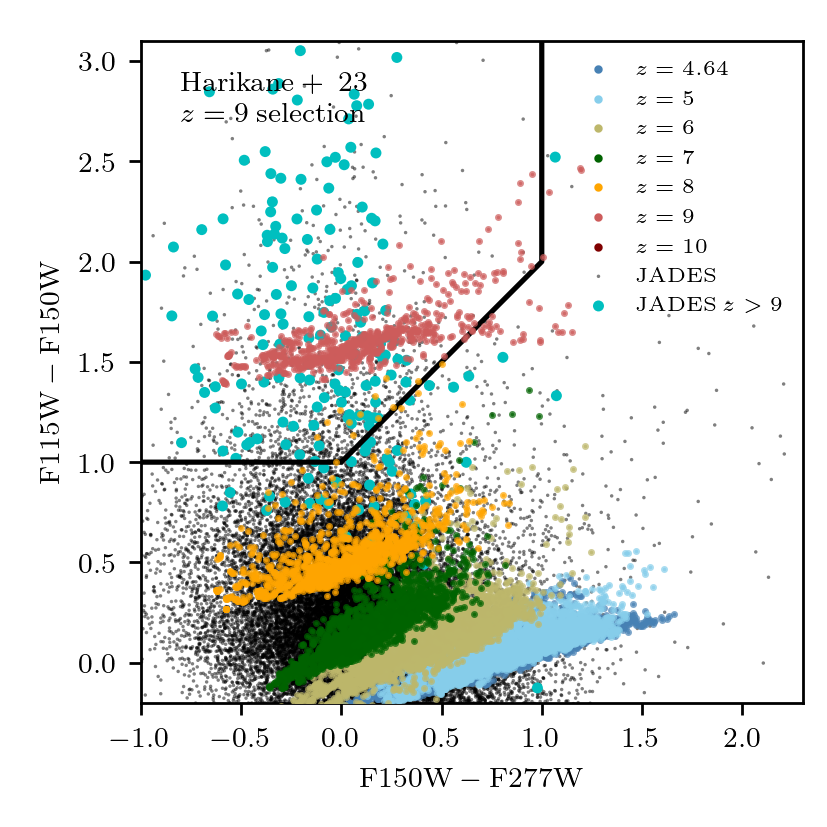}
\caption{Example colour-colour diagram of JWST filters F115W$-$F150W versus F150W$-$F277W. \sphinx{} galaxies are shown as coloured data points. Small black points represent galaxies from JADES and cyan points are JADES galaxies with photometric or spectroscopic redshifts of $z>9$. For comparison, the region marked in black shows the $z=9$ selection criteria from \protect\cite{Harikane2023}. Note that all \sphinx{} galaxies at $z=10$ are above the y-range of the plot.}
\label{fig:colour-colour}
\end{figure}

\subsubsection{Photometric redshifts}
\begin{figure}
\includegraphics[width=0.45\textwidth]{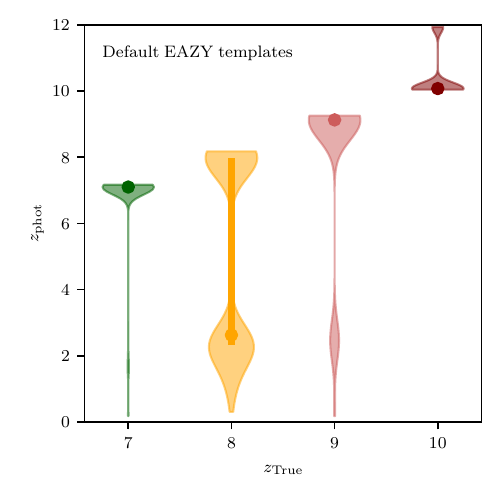}
\includegraphics[width=0.45\textwidth]{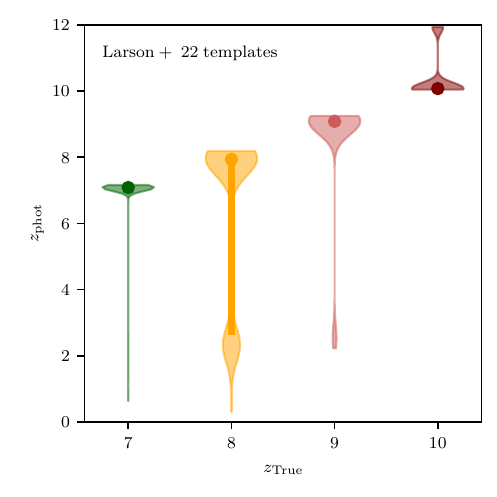}
\caption{Violin plots of photometric redshifts estimated with EAZY using the default templates (top) or the templates from \protect\cite{Larson2022} (bottom) versus the true redshift of the galaxies in the \sphinx{} database at $z\geq7$. The data points represent the median redshift and the vertical line represents the interquartile range (which is only visible for the $z=8$ distribution).}
\label{fig:photo-zs}
\end{figure}

Beyond colour-colour diagrams, one can attempt to estimate the exact redshifts of the galaxies from their photometry. As an example, in Figure~\ref{fig:photo-zs}, we show a violin plot of the distribution of photometric redshifts estimated with {\small EAZY} \citep{Brammer2008} using all of the wide and medium band filters in JWST for all galaxies in the \sphinx{} data release at $z\geq7$. The top panel shows the results using the default EAZY FSPS templates while the bottom panel uses the templates designed for high-redshift galaxies from \cite{Larson2022}. For the fits, we have assumed a 5\% error on the flux in each filter. For galaxies at $z=7$ and $z=9$, the vast majority of galaxies are assigned the correct redshift, with only a few galaxies being placed at low-redshift. At $z=10$, the photometric redshift estimation is also very good with a few systems preferring a higher redshift solution which is the upper limit of our prior. Interestingly, when the default templates are used, most of the galaxies in the database at $z=8$ are assigned a low-redshift solution where the Lyman break is confused with a Balmer break. This is likely due to the fact that the Lyman break occurs towards the edge of the F115W filter so the bluest band has much lower flux than the second bluest (F140M), thus mimicking a weak break (as can be seen in Figure~\ref{fig:colour-colour}). Hence, by working with only the maximum likelihood redshifts, there could be a significant population of $z=8$ galaxies that are incorrectly assigned a low-redshift solution in real JWST data. This confusion does decrease significantly when adopting the templates from \cite{Larson2022}. Here we have used every JWST filter; however, the vast majority of imaging surveys use far fewer filters, which would make photometric redshift estimation even more uncertain than in this experiment.

\newpage
\subsubsection{UV spectral slopes} \label{sec:uv_slopes} Prior to JWST, UV spectral slope, $\beta$, represented one of the few diagnostics available that could be directly measured from observations \citep[e.g.][]{Dunlop2012,Finkelstein2012,Bouwens2014,Wilkins2016a,Bhatawdekar2021}. Because $\beta$ depends on both age and metallicity of the underlying stellar population as well as the dust content in galaxies, it traces the recent SFH of a galaxy. Particular emphasis is placed on finding galaxies with extremely blue UV slopes, which would be indicative of very metal-poor or even metal-free stellar populations \citep[e.g.][]{Topping2022}. In principle, $\beta$ can be measured for large samples of galaxies from photometry, making it one of the primary constraints from early JWST data.

\begin{figure}
\includegraphics[width=0.45\textwidth]{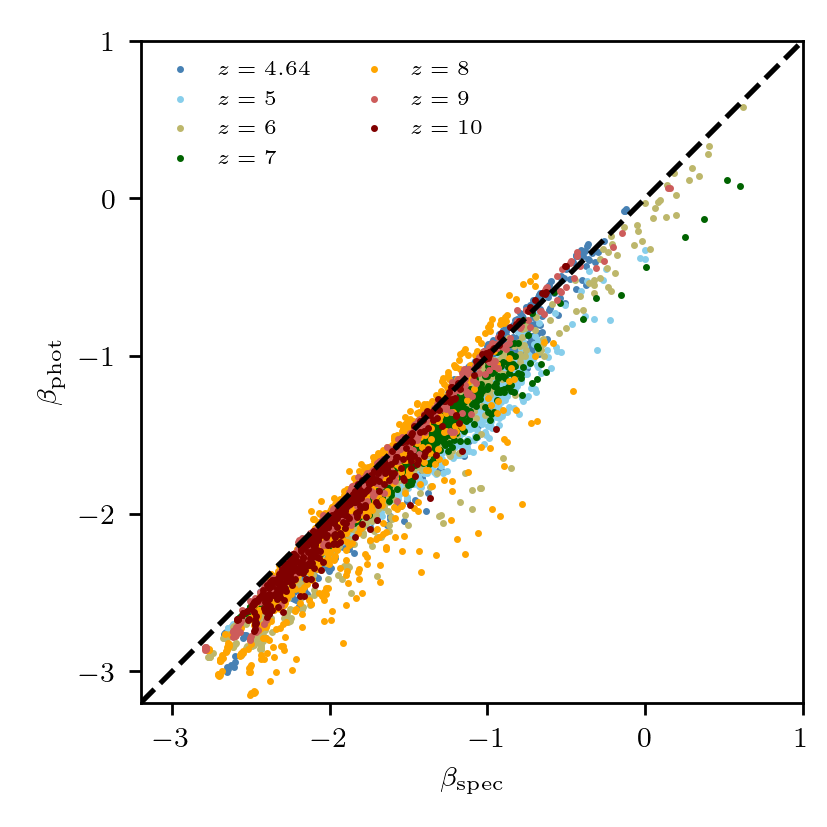}
\caption{Comparison between the UV slope measured photometrically and that from the spectra for \sphinx{} galaxies at various redshifts. The dashed black line represents the one-to-one relation. In general, UV slopes measured from photometry tend to be steeper than slopes measured from the spectra.}
\label{fig:bsbp}
\end{figure}

Using both the photometry and spectra in \spdrone, in Figure~\ref{fig:bsbp} we show $\beta$ measured from the photometry for each \sphinx{} galaxy along each sight line versus the true continuum slope along that direction (i.e. the stellar $+$ nebular continuum attenuated by dust). In general, most of the points fall close to the one-to-one relation (dashed black line) and the median absolute difference is 0.12 with a bias such that photometrically measured $\beta$ is bluer than that from the spectra. There are two effects that can cause this behaviour. Emission lines are present in the observed photometry and can thus bias the beta measurement. To assess the importance of emission lines, we have generated the photometry for each galaxy without emission lines and fit the UV slope. In doing this, we find that the median absolute difference drops to 0.08 and thus emission lines are not the dominant source of error. The remaining and dominant source of error is thus due to the shape of the filter response curves. Both biases can partially be removed with SED fitting codes. Nevertheless, naive measurements of $\beta$ that do not correct for these effects can exhibit a typical bias of 0.12. Our measurement was made for galaxies using all of the wide and medium band filters on NIRCam and results may be different for different subsets of filters. We have also tested using only the wide filters and find similar results. Furthermore, we have assumed an IGM that fully attenuates Ly$\alpha$ and the bias may be stronger for Ly$\alpha$-emitting galaxies that live in ionized bubbles.

\begin{figure}
\includegraphics[width=0.45\textwidth]{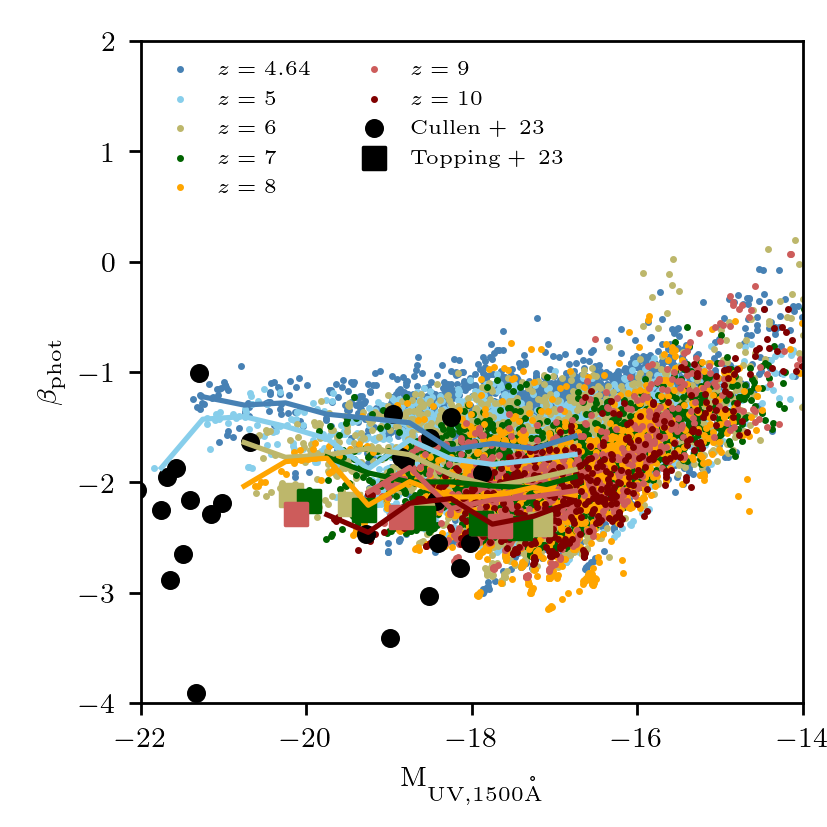}
\caption{$\beta$ measured photometrically versus ${\rm M_{UV}}$ for \sphinx{} galaxies compared to individual high-redshift JWST measurements from \protect\citep{Cullen2023} as well as mean relations of high-redshift JADES galaxies from \protect\citep{Topping2023}. The colour of the points indicates the redshift and the solid lines represent the binned median values.}
\label{fig:beta_muv}
\end{figure}

The UV spectral slope is often compared to the UV magnitude of a galaxy. In Figure~\ref{fig:beta_muv} we compare \sphinx{} galaxies to high-redshift photometrically-selected JWST $\beta$ measurements from \cite{Cullen2023,Topping2023}. The \cite{Cullen2023} measurements represent individual galaxies that exhibit significant scatter, often to unphysically blue values of $\beta$. In the magnitude regime where the measurements overlap, the scatter is consistent with what we see in \sphinx{}. The points from \cite{Topping2023} represent the mean relation for samples of galaxies photometrically selected at different redshifts. At the faint-end, there is little observed evolution in $\beta$ between $z=6$ and $z=9$, consistent with \sphinx{}. The typical JADES galaxy in \cite{Topping2023} is slightly bluer than what we find in \sphinx{}, consistent with our earlier comparison to JADES galaxies. Among the many effects that can cause this are the choice of dust attenuation curve in \sphinx{} and the flux limit constraints in JADES (i.e. redder galaxies at the faint-end are less likely to meet signal-to-noise thresholds). 

At UV magnitudes fainter than $-17$ there is an upturn in the $\beta-{\rm M_{UV}}$ relation in \sphinx{}. This is due to the SFR threshold used to create \spdrone\ and would go away if we include the more numerous and fainter galaxies with ${\rm SFR}<0.3\ {\rm M_{\odot}\ yr^{-1}}$.

\subsubsection{UV luminosity function}
\begin{figure}
\includegraphics[width=0.45\textwidth]{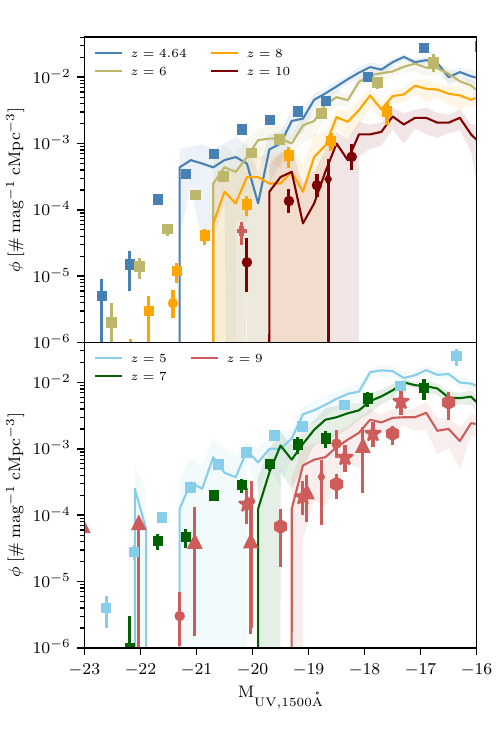}
\caption{UV luminosity function for \sphinx{} galaxies at different redshifts compared to high-redshift photometric constraints. The curves show the mean UV luminosity functions over the ten viewing angles and the shaded areas correspond to the standard deviation. Squares are observational data from HST at $z\leq8$ \protect\cite{Bouwens2021}, while circles, triangles, diamonds, stars, cross, and hexagons show JWST measurements at $z\geq8$ from \protect\cite{Donnan2023}, \protect\cite{Harikane2023} and \protect\cite{Bouwens2023}, \protect\cite{Leung2023}, \protect\cite{Franco2023}, respectively. The cutoff in \sphinx{} data at high bright magnitudes is due to the finite volume of our simulation box while the downturn at faint magnitudes is a result of our SFR threshold.}
\label{fig:uv_lum_func}
\end{figure}

One of the earliest results from JWST was constraints on the high-redshift UV luminosity function \cite{Bouwens2023,Harikane2023,Donnan2023}.  Results for the \sphinx{} UV luminosity function (dust-attenuated from an angle-averaged perspective) were compared to HST data in \cite{Rosdahl2022}. The \sphinx{} data release now contains the UV magnitudes along ten different lines of sight. In Figure~\ref{fig:uv_lum_func} we show that the luminosity functions in the \sphinx{} data release (averaged over the ten directions) are in good agreement with both photometric constraints from HST and JWST from $z\approx 5$ to 10. The overlap between observations and data at $z=9$ and 10 only spans $\sim2$ magnitudes; however, this will improve as observations go deeper and gravitational lensing probes fainter galaxies. We note that the turnover in our luminosity function at ${\rm M_{ UV,1500\AA}} \approx -17$ is due to our SFR threshold and not due to dwarf galaxy quenching from reionization or any other process. It is clear that dust attenuation remains a necessary ingredient for agreement with observations and the UV luminosity function is strongly sensitive to the chosen dust model. However, we stress that our assumed dust model was in no way designed to guarantee such an agreement and perhaps even makes faint galaxies too red (see above discussion).

\subsubsection{${\rm M_*}$ - $\MUV$}

\begin{figure}
\centering
\includegraphics[width=0.45\textwidth]{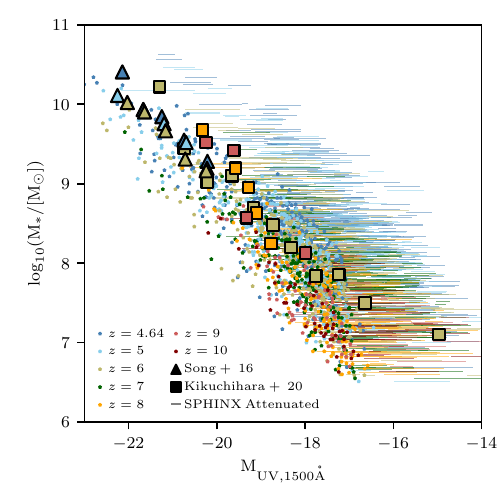}
\includegraphics[width=0.45\textwidth]{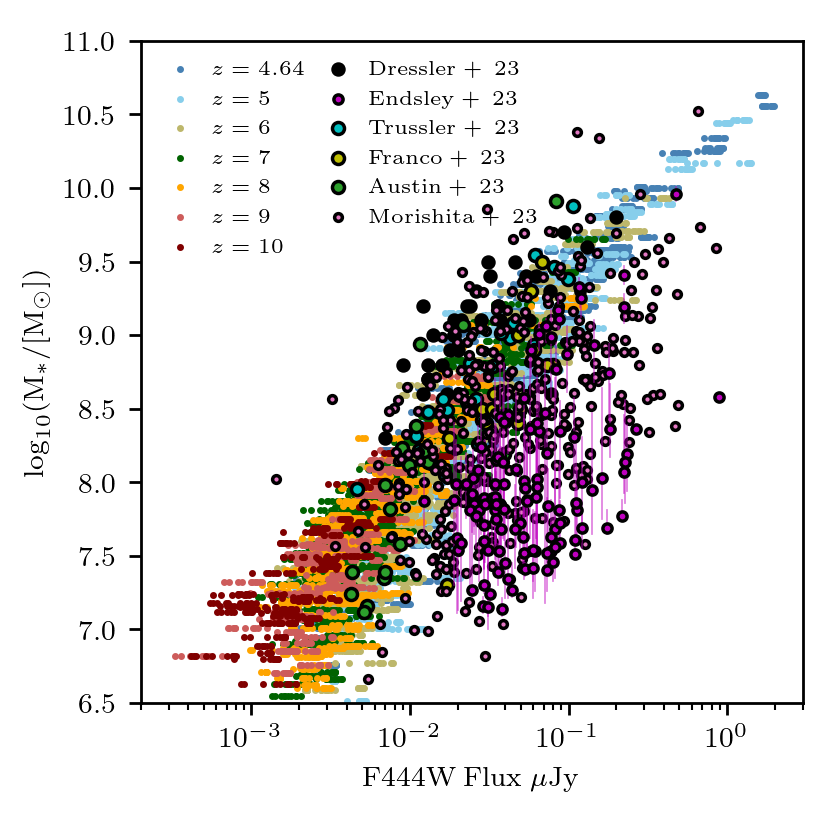
}
  \caption{(Top) Galaxy stellar mass as a function of UV magnitude at 1500~\AA{} at various redshifts. The points represent intrinsic magnitudes for \sphinx{} galaxies, while horizontal lines represent the range of attenuated magnitudes along the 10 directions. For comparison we show observational inferences from HST $+$ Spitzer data \protect\citep{Song2016,Kikuchihara2020}. (Bottom) Galaxy stellar mass as a function of F444W flux in $\mu$Jy. Over-plotted are stellar masses inferred from JWST by CEERS \protect\citep[magenta, $z\sim7-8$,][]{Endsley2023}, GLASS \protect\citep[black, $z\sim5-7$,][]{Dressler2023}, PEARLS \protect\citep[cyan, $z\sim7-12$,][]{Trussler2023}, COSMOS-Web \protect\citep[yellow, $z\sim9-11$,][]{Franco2023}, NGDEEP \protect\citep[green, $z\sim8-16$][]{Austin2023}, and a compilation from \protect\cite{Morishita2023} at $z=5-13$. }
  \label{fig:MUV_vs_mstar}
\end{figure}

In Section~\ref{sec:sfh}, we connected UV magnitudes ($\MUV$) with SFR. Since SFR and stellar mass are correlated via the star formation main sequence, $\MUV$ also correlates with stellar mass. In the top panel of Figure~\ref{fig:MUV_vs_mstar} we show stellar mass as a function of  intrinsic and attenuated UV magnitudes at 1500~\angstrom{} for galaxies at different redshifts in \spdrone. We show  intrinsic magnitudes as dots and in horizontal lines the range of attenuated magnitudes along the ten different sight lines. We also show fits to observational inferences from \cite{Song2016} and \cite{Kikuchihara2020}. The \sphinx{} galaxies compare well against these observations only if dust attenuation is ignored. At face value, this can be interpreted as \sphinx{} galaxies being too massive and/or having too strong dust attenuation for a fixed UV magnitude, both plausible outcomes if star formation is happening too rapidly very early in the evolution of our simulated galaxies. This could also be caused by an attenuation curve that is too strong at rest-frame UV wavelengths or any of the other possibilities discussed in Section~\ref{sec:phot}. However, we emphasize that there remains significant uncertainty on the accuracy of stellar mass estimated from photometry.

\subsubsection{${\rm M_*}$ - ${\rm F444W}$}
Rather than compare stellar masses with UV magnitude, one can alternatively compare stellar mass with F444W flux. The advantages of doing so are 1) F444W flux is much less sensitive to dust than UV magnitude so using this filter avoids some of the uncertainty in dust modelling for high-redshift galaxies, 2) it samples a longer wavelength part of the spectrum which is more sensitive to radiation from older stellar populations than UV magnitude and should thus better probe stellar mass, and 3) it is a directly observable quantity and therefore does not require modelling. Hence the scatter between stellar mass and observed F444W flux at fixed redshift should be lower than between stellar mass and UV magnitude.

\begin{figure}
\centering
\includegraphics[width=0.45\textwidth]{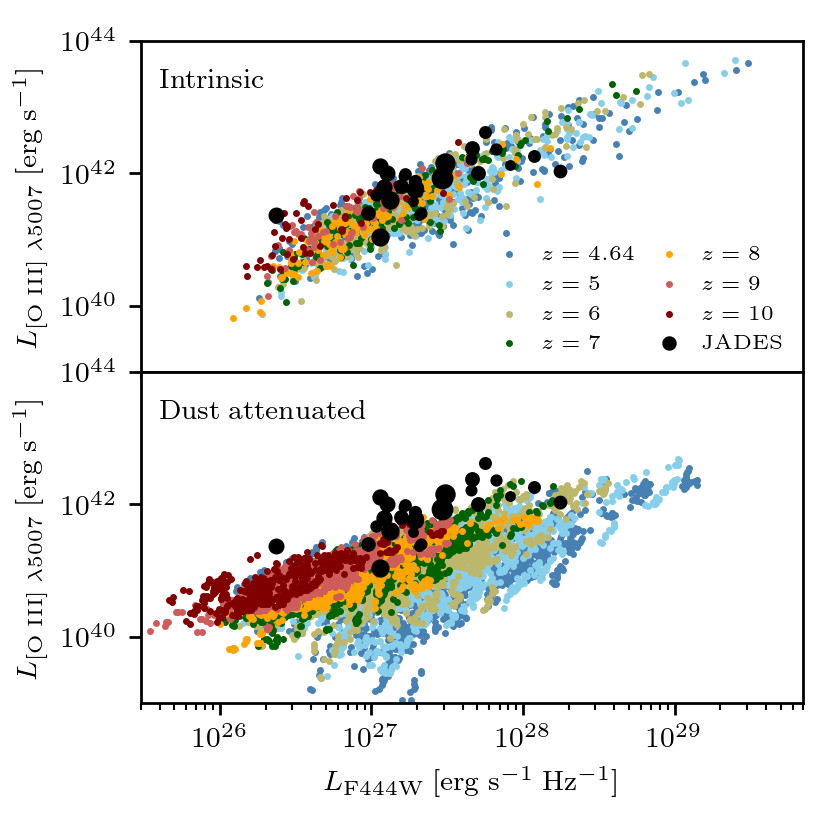}
  \caption{[O~{\small III}]~$\lambda$5007 luminosity as a function of F444W luminosity for \sphinx{} galaxies. We show both the intrinsic values (top) and dust attenuated (bottom). Black points represent high-redshift galaxies from JADES where the size of the point is indicative of redshift, with larger points being at higher redshift.}
  \label{fig:F444W_OIII}
\end{figure}

In the bottom panel Figure~\ref{fig:MUV_vs_mstar} we plot stellar mass as a function of flux in the NIRCam F444W filter for \sphinx{} galaxies as well as a large collection of observational inferences of stellar mass as a function of F444W from multiple JWST programs: CEERS \citep[magenta, $z\sim7-8$,][]{Endsley2023}, GLASS \citep[black, $z\sim5-7$,][]{Dressler2023}, PEARLS \citep[cyan, $z\sim7-12$,][]{Trussler2023}, COSMOS-Web \citep[yellow, $z\sim9-11$,][]{Franco2023}, NGDEEP \citep[green, $z\sim8-16$,][]{Austin2023}, and a compilation from \cite{Morishita2023} at $z=5-13$. These programs all sample similar redshifts as the \sphinx{} data. However, despite \sphinx{} galaxies showing a rather tight trend, there is significant scatter in the data reported from the literature. 

There are numerous physical effects that can drive scatter in this relation. 1) As redshift increases, the F444W filter samples a different part of the galaxy SED, 2) As redshift increases, so does the luminosity distance, which will decrease the observed flux, and 3) the [O~{\small III}]~$\lambda$5007$+$H$\beta$ emission lines enter the F444W filter at redshifts slightly below $z=7$. Despite these effects, \sphinx{} galaxies still exhibit a rather tight trend; although, there is clear evidence of the redshift effect. Nevertheless, these effects cannot completely explain what is reported in the literature. Most notably, the \cite{Dressler2023} sample is at lower redshift than the \cite{Endsley2023} sample and predicts much higher stellar masses, yet both exhibit similar ranges of F444W fluxes. Hence the \cite{Dressler2023} galaxies should have lower stellar masses than those in \cite{Endsley2023}, unless [O~{\small III}]~$\lambda$5007$+$H$\beta$ overwhelmingly dominate the filter in the \cite{Endsley2023} sample. This is unlikely given how wide the filter is and the discrepancy in stellar mass is up to three orders of magnitude. In Figure~\ref{fig:F444W_OIII} we show the F444W flux against [O~{\small III}]~$\lambda$5007 luminosity for \sphinx{} galaxies, both intrinsic (top) and dust attenuated (bottom). The JADES galaxies have a typical [O~{\small III}]~$\lambda$5007 luminosity of $10^{42}\ {\rm erg\ s^{-1}}$ and can thus increase the flux in the band by around a factor of two. This is clearly not enough to explain the stellar mass discrepancy. However, consistent with our above JADES discussion, this comparison shows that the dust model in \sphinx{} is likely too strong at all wavelengths. Removing dust would shift the \sphinx{} galaxies to the right on the stellar mass-F444W relation.

In contrast to the \cite{Dressler2023} and \cite{Endsley2023} inferences, the results from \cite{Trussler2023,Austin2023,Franco2023} all overlap with \sphinx{} data. We can speculate the reason for such different stellar mass estimates is somehow related to the way the SEDs are being fit, the choice of emission line library and the prior on SFH. However, based on these results, we emphasize that before they can be used on constraints for galaxy formation physics, stellar mass estimates and therefore any relation involving stellar mass (e.g the stellar mass-UV magnitude relation, stellar mass-halo mass relation) should be made consistent amongst the various observational campaigns.

\subsubsection{Galaxy sizes}
Galaxy sizes are computed in a similar manner to what is typically done for real observations. For each mock observation in \spdrone\ we sum the stellar continuum, nebular continuum, and nebular emission line images for each JWST filter and use the {\small PHOTUTILS} \citep{photutils} package to segment the image. We select the segment corresponding to the central galaxy in each halo and measure the flux of the segment\footnote{Note that this flux of the main segment can differ significantly from the flux of all stars within the virial radius of the halo. This is particularly true in the case of a merger.} as well as the circularized half-light radius. By doing this, in principle, we can directly compare galaxy sizes with observations. 

\begin{figure}
\includegraphics[width=0.45\textwidth]{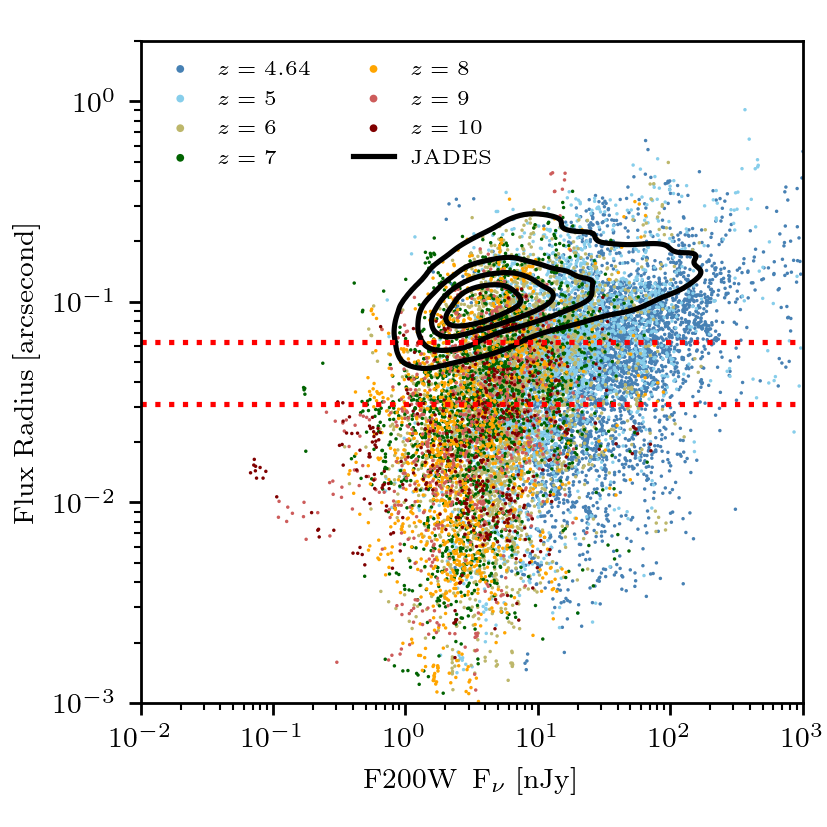}
\caption{Circularized half-light radii in arcseconds for \sphinx{} galaxies in the NIRCam F200W filter as a function of their segment flux in nJy. For comparison, we show JADES galaxies with $z_{\rm phot}>4$ and an ${\rm S/N>3}$ in the F200W filter as black contours. The dotted horizontal red lines show the pixel sizes of the short and long wavelength channels of NIRCam. Galaxies smaller than this would appear as point sources convolved with the filter PSF.}
\label{fig:size-photutils}
\end{figure}

In Figure~\ref{fig:size-photutils} we show the circularized half-light radius in arcseconds as a function of the segment flux in nJy in the F150W filter. For comparison, we show recent data from JADES \citep{Bunker2023b,Eisenstein2023,Hainline2023,Rieke2023} for galaxies that have a photometric redshift of $z_{\rm phot}>4$. There is significant overlap between the observational and simulated data sets in terms of segment fluxes; however, \sphinx{} galaxies scatter to significantly smaller radii than what is observed with JWST. This is because we have not convolved our images with a PSF. The horizontal dotted red lines show the pixel sizes for the short and long-wavelength modes of NIRCam and thus all \sphinx{} galaxies with a radius smaller than this line would likely be observed as a point source with a size given by the PSF for the particular filter. The sizes of resolved sources in our data set are very consistent with what is observed in JADES.

For individual galaxies, we find a significant amount of dispersion in their measured sizes depending on viewing angle. One can interpret that a significant fraction of the observed dispersion in observations is explained by viewing geometry, without needing to invoke intrinsic dispersion in SFR, star formation history or extinction at a given magnitude, although variations in all these parameters should be expected. In certain cases, the galaxy size can vary by more than an order of magnitude. Part of this is due to the fact that the galaxies are very clumpy and frequently undergo mergers. If the merger is aligned along the line of sight, then the galaxy will have a very small effective size while if the merger is perpendicular to the sight line, the flux radius becomes very extended. The problem becomes exacerbated after convolution with a PSF because these galaxies are often small enough where the merger becomes blended.
\vspace{1cm}
\subsubsection{Size - luminosity relation}

\begin{figure}
\includegraphics[width=0.45\textwidth]{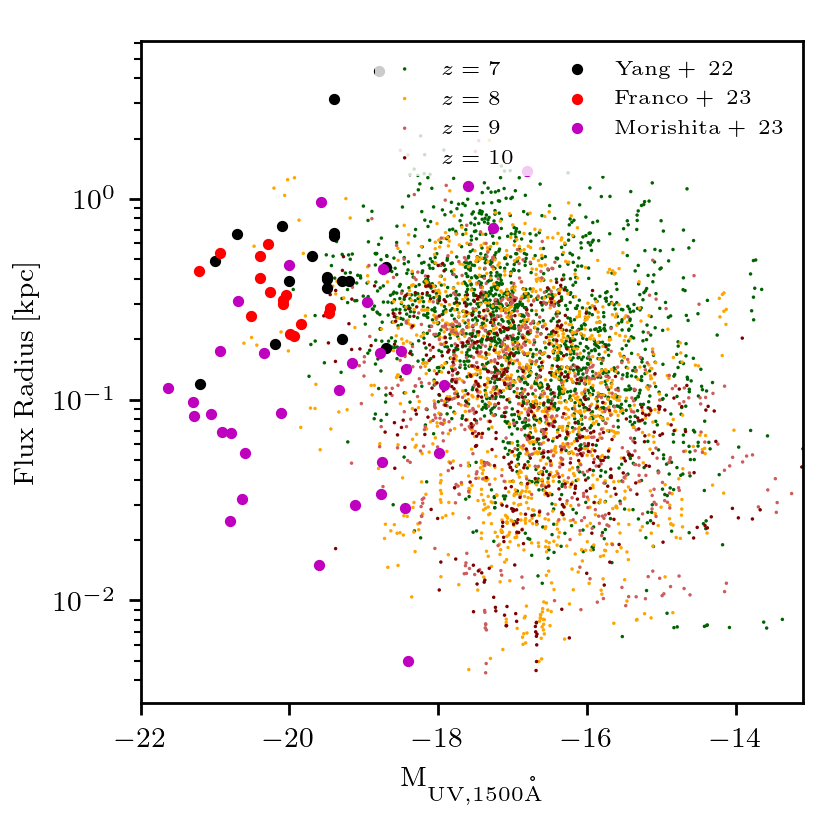}
\caption{Size-luminosity relation for \sphinx{} galaxies at $z\geq7$ in the JWST F277W filter. For comparison, we show recent JWST observations in the same filter from \protect\cite{yang2022_size_lum,Franco2023} as black and red points, respectively, as well as various high-redshift dropout selected galaxies from \protect\cite{Morishita2023} in magenta.}
\label{fig:size-lum}
\end{figure}

While in the previous section we compared galaxy sizes with their observed flux density, galaxy size is more commonly compared to the intrinsic galaxy UV luminosity \citep[e.g.][]{Kawamata2018,Bouwens2022size}. In Figure~\ref{fig:size-lum} we show galaxy size in the JWST F277W filter against the observed UV magnitude for \sphinx{} galaxies at $z\geq7$ compared with JWST observations in the same filter and redshift interval from \cite{yang2022_size_lum,Franco2023}. In general, fainter galaxies have smaller sizes, but we predict a significant $>1$~dex scatter at any magnitude. There are simply not enough observed galaxies that overlap in magnitude with \spdrone\ to make any assessment on the agreement between observations and our simulations\footnote{A further analysis of JADES public data would likely have much more overlap due to the deeper observations.}. Furthermore, we highlight that despite our attempt to measure sizes in a similar manner to observers, differences remain, beyond the fact that we have not convolved with the PSF and added noise. For example, it is typical to fit observations with a Sérsic profile with a slope of $n=1$. Empirically, we find that this does not represent the clumpy nature of our galaxies. Thus the reported definition of radius can vary between different observational studies and what we have measured for \sphinx{} galaxies. Nevertheless, because we provide the raw images with the data release, it is possible to create a one-to-one mapping by applying the same methods to both observations and the \sphinx{} galaxies.

\subsection{Spectra}
One of the key advantages of JWST in comparison to previous instruments is its spectroscopic capabilities at high-redshift. In this section, we both describe the spectral data products and demonstrate a few example use-cases of the \sphinx{} public data release.

The spectra are stored at high resolution (1~\AA) and can be degraded to match any instrument with the appropriate line-spread function.

\subsubsection{E(B$-$V)}

\begin{figure}
\includegraphics[width=0.45\textwidth]{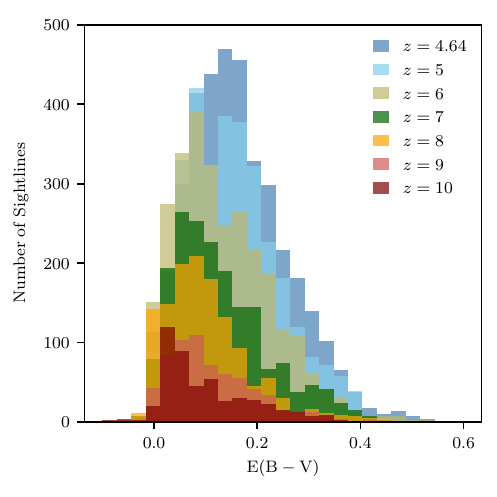}
\caption{E(B$-$V) distributions for spectra in the \sphinx{} database as a function of redshift.}
\label{fig:ebmv}
\end{figure}

As galaxies chemically evolve so does their dust content which can have a significant impact on the observed spectrum. We show the impact of dust in the form if interstellar reddening, E(B$-$V), as measured from the Balmer decrement (H$\alpha$/H$\beta$) in Figure~\ref{fig:ebmv}. As redshift decreases, the typical E(B$-$V) of star-forming galaxies increases from $<0.1$ at $z\geq9$ to nearly 0.2 at $z=4.64$. Since the database contains galaxies with both high and low reddening, it can be used, for example, to test our ability to dust-correct spectra and to predict completeness rates for luminosity functions.

\subsubsection{UV spectral slopes} 
\label{sec:age_vs_beta}
\begin{figure}
  \centering
  \includegraphics[width=0.45\textwidth,trim={0 1.4cm 0 0.7cm},clip]{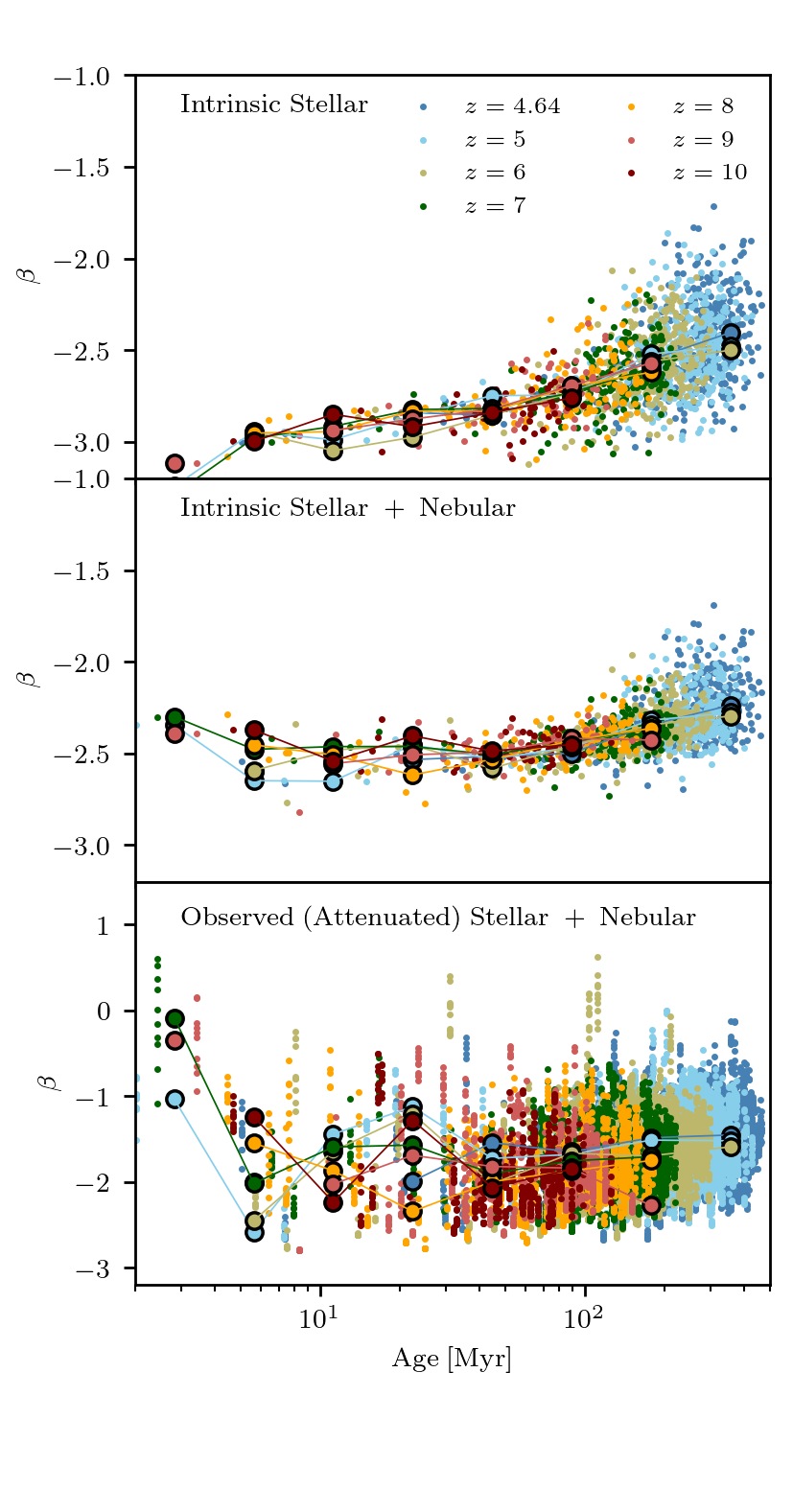}
  \caption
  {UV slope, $\beta$, as a function of mass-weighted galaxy age, measured from the intrinsic stellar continuum (top), intrinsic stellar + nebular continuum (middle), and dust attenuated stellar + nebular continuum along each of the ten sight lines (bottom) for each \sphinx{} galaxy. Small circles represent individual galaxies and the large connected circles represent mean values per age bin. Colour indicates the redshift of the galaxy.}
  \label{fig:beta_vs_age}
\end{figure}

We return to UV spectral slopes to demonstrate how the different components of the spectrum (stellar continuum, nebular continuum, and dust) impact the observed $\beta$ of a galaxy. We show in Figure~\ref{fig:beta_vs_age} the spectral slope versus galaxy age. The slope is computed between 1400~\AA{} and 2500~\AA{} (rest-frame) and the galaxy age is the mass-weighted mean age of all star particles in the galaxy. In the top panel, $\beta$ is computed from only the intrinsic stellar continuum. Here we find a clear correlation between age and $\beta$, with $\beta \sim -3$ for the youngest galaxies and $\sim -2.5$ for the oldest ones. This highlights the expected intrinsic distribution of UV slopes given the complex SFHs in \sphinx{}. The middle panel shows the same, but now with added contributions from the nebular continuum. This modifies the correlation between $\beta$ and age, with the nebular continuum significantly reddening the youngest galaxies. When the nebular continuum is added, there is indeed very little correlation between the mean mass-weighted age of the galaxy and $\beta$. This is a combination of two effects. First, the nebular continuum preferentially reddens the youngest and bluest galaxies. Second, even for galaxies with relatively older ages, the UV part of the spectrum is still often dominated by the youngest stars which can easily out-shine the older population. This conspires to produce a very flat trend of $\beta$ with age.

However, when dust is included, the intrinsic trend between $\beta$ and age can be further altered. In the bottom panel, we compute $\beta$ from the stellar plus nebular continuua, dust attenuated along each of the ten sight lines. In this case, the slope is widely scattered to very large $\beta$ values, even positive ones, and no correlation between $\beta$ and galaxy age remains. It is therefore impossible to determine anything about the ages of stellar populations in galaxies from observed UV slopes, unless a correction for dust attenuation and nebular continuum can be accurately made. Many modern SED fitting codes account for these effects simultaneously and we anticipate that the \sphinx{} data can be used to test the accuracy of such methods.

\subsubsection{Emission line luminosity functions}
\begin{figure}
\includegraphics[width=0.45\textwidth]{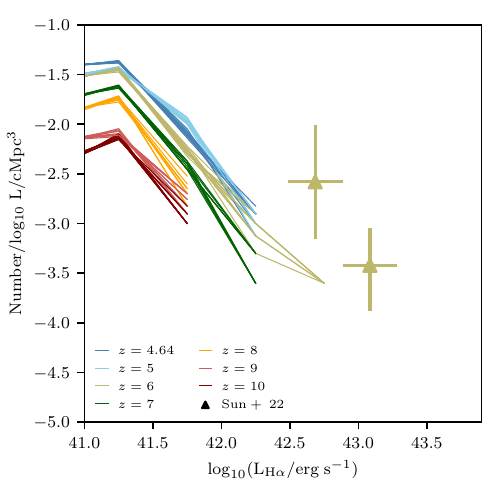}
\includegraphics[width=0.45\textwidth]{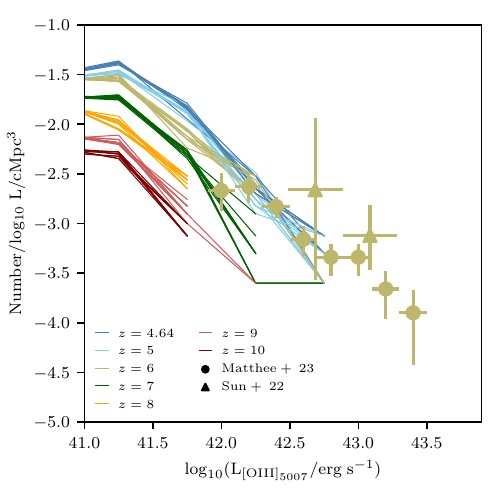}
\caption{Line luminosity functions for H$\alpha$ (top) and [O~III]~$\lambda$5007 (bottom) as a function of redshift. For each redshift, we show a luminosity function for each of the ten different sight lines to demonstrate the expected variance with viewing angle. For comparison we show results from $z\sim6$ JWST observations from \protect\cite{Matthee2023,Sun2022}.}
\label{fig:llf}
\end{figure}

Because \sphinx{} is a full-box simulation with initial conditions selected to produce a typical halo mass function at $z=6$, it can be used to predict emission line luminosity functions in the same way that we predicted the UV luminosity function. In Figure~\ref{fig:llf} we show reddening corrected H$\alpha$ (top) and [O~III]~$\lambda$5007 (bottom) luminosity functions at multiple redshifts compared to JWST observational constraints from \cite{Matthee2023,Sun2022}. For each redshift, we show ten luminosity functions representing each of the ten mock sight lines. This demonstrates the sample variance in the \sphinx{} data set. As redshift decreases, the number density of line emitters increases. We note that the turnover at the faint-end of the luminosity function is due to our SFR cut rather than a physical effect, for example, suppression due to the radiative feedback from reionization. 

The \sphinx{} [O~III]~$\lambda$5007 luminosity function is in very good agreement with results from the Eiger survey \citep{Kashino2023,Matthee2023} but under-predicts  estimates from the four serendipitous line emitters detected in JWST commissioning data \citep{Sun2022}. This likely explains why the \sphinx{} H$\alpha$ luminosity function under-predicts the same commissioning data. We note that the luminosity function from the commissioning data is based on only four galaxies.

\subsubsection{Halo mass - Line luminosity relations}

Understanding how halo mass connects with emission line luminosities is not only important for understanding the galaxy-halo connection at high redshift, but it is a key input for interpreting intensity mapping surveys. For example, SPHEREx \citep{Dore2014}, CONCERTO \citep{CONCERTO2020}, and TIME \citep{Crites2014} can conduct high-redshift intensity mapping for emission lines in the rest-frame UV, optical, and IR up to high redshift. 

\begin{figure}
\includegraphics[width=0.45\textwidth,trim={0 0.5cm 0 0.5cm},clip]{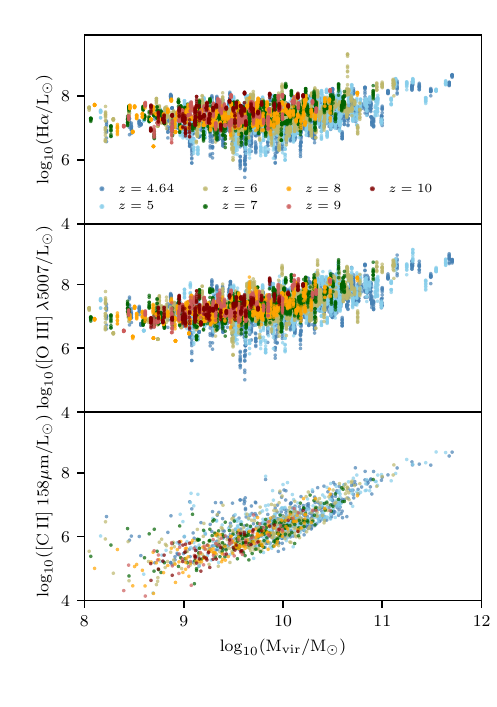}
\caption{Observed H$\alpha$ (top), [O~{\small III}]~$\lambda$5007 (middle), and [C~{\small II}]~158$\mu$m (bottom) emission versus halo virial mass. Because [C~{\small II}]~158$\mu$m is not impacted by dust, we show only the intrinsic value of the emission line as it is the same across all ten sight lines.}
\label{fig:mvir-eline}
\end{figure}

In Figure~\ref{fig:mvir-eline} we show the relationship between observed H$\alpha$, [O~{\small III}]~$\lambda$5007, and [C~{\small II}]~158$\mu$m emission and halo virial mass. While there is significant scatter in the relationship between nebular lines and halo mass, there is a much tighter correlation between [C~{\small II}]~158$\mu$m and halo mass as this line tends to trace neutral gas. The \sphinx{} public data release can be used to build relations between any of the available emission lines and intrinsic galaxy properties which can be used to populate larger simulations and make predictions for upcoming intensity mapping experiments. 

\subsubsection{Equivalent width distributions}
\begin{figure}
\includegraphics[width=0.45\textwidth]{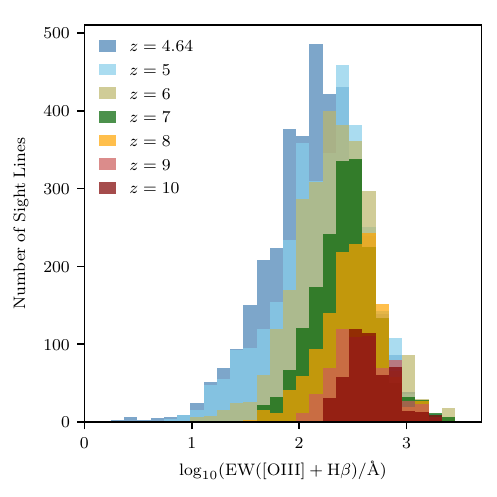}
\caption{Equivalent width distributions of [O~{\small III}]~$\lambda$4959,5007$+$H$\beta$ for \sphinx{} galaxies at each redshift.}
\label{fig:o3hb_ews}
\end{figure}

Because the \sphinx{} data set contains both emission line luminosities and the stellar (and nebular) continuum, it is trivial to compute equivalent widths (EWs) for any emission line in our sample. In \Fig{fig:o3hb_ews} we show a histogram of [O~{\small III}]~$\lambda$4959,5007$+$H$\beta$ EWs for each redshift bin in the data set. As redshift increases, the typical EW in our data set also increases due to the fact that we are sampling galaxies further above the main-sequence. One important feature is that the data release contains nearly 400 sight lines with ${\rm EW}>1000 \ \text{\AA}$, which are typically rare in simulations. The \sphinx{} data release is among the first along with the FLARES simulation \cite{Wilkins2023b} to produce galaxies with such extreme emission lines. However, neither simulation produces the long tail of high EW galaxies observed at high-redshift \citep[e.g.][]{debarros2019}. The origin of this discrepancy is currently unknown.

\subsubsection{Diagnostic diagrams}
Diagnostic diagrams, for example the BPT and VO diagrams \citep{Baldwin1981,Veilleux1987}, are important tools for understanding the excitation mechanisms and conditions of the ISM throughout the Universe. Understanding how these diagrams are expected to evolve with changing ISM conditions as a function of redshift is key for interpreting JWST spectra. For example, it is now well established that high-redshift galaxies are offset on the BPT diagram compared to their low-redshift counterparts \citep[e.g.][]{Steidel2014} and using the BPT to identify AGN at high-redshift may fail \citep[e.g.][]{Groves2006,Feltre2016}.

\begin{figure}
\includegraphics[width=0.45\textwidth]{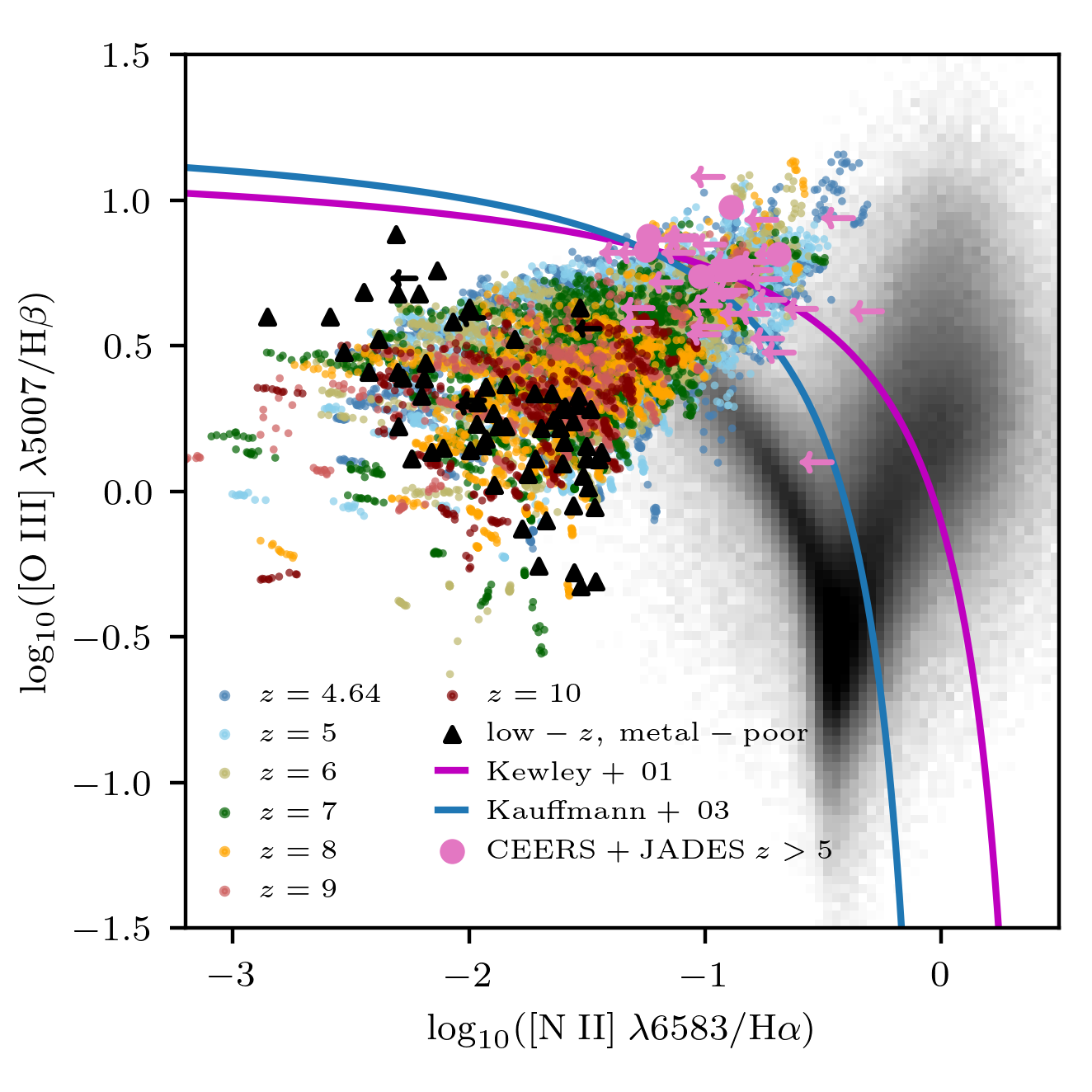}
\includegraphics[width=0.45\textwidth]{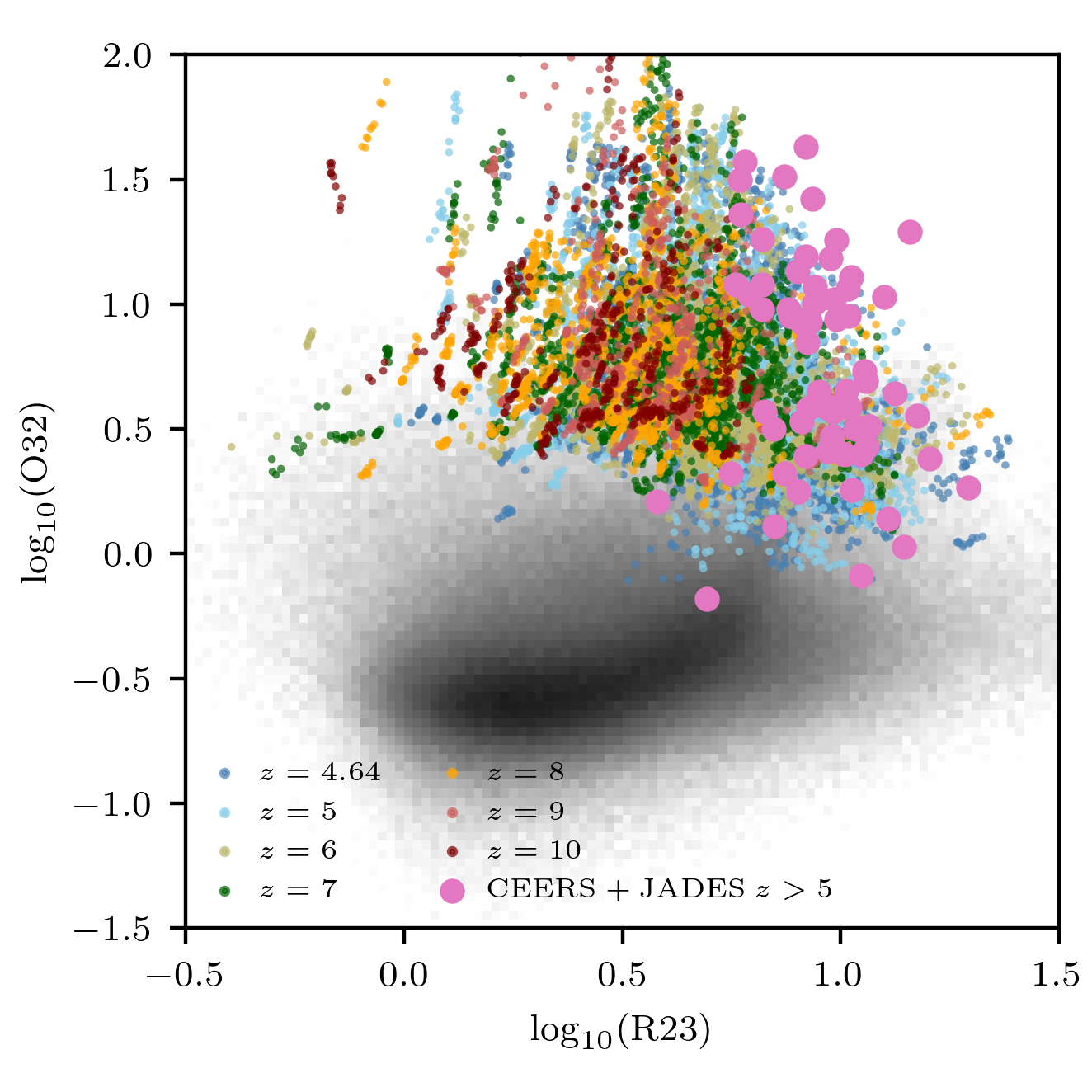}
\caption{(Top) N~{\small II} BPT diagram for \sphinx{} galaxies compared to SDSS \protect\citep[grey 2D histogram,][]{Thomas2013}, local metal-poor galaxies \protect\citep[black triangles,][]{Isobe2022,Guseva2017}, and high-redshift JWST galaxies \protect\citep[pink circles,][]{Sanders2023,Cameron2023}. (Bottom) R23-O32 diagram for \sphinx{} galaxies compared to SDSS  and high-redshift JWST galaxies.}
\label{fig:bpt}
\end{figure}

In Figure~\ref{fig:bpt} we show an example BPT diagram of \sphinx{} galaxies using the dust-attenuated and de-reddened emission line luminosities along the ten sight lines. The data set nicely maps out the region populated by the lowest metallicity galaxies known in the low-redshift Universe \citep[black triangles,][]{Isobe2022,Guseva2017} as well as JWST observations of $z>5$ star-forming galaxies from CEERS and JADES \cite[pink circles,][]{Sanders2023,Cameron2023}.

The upper cutoff in our diagram is driven by the fact that we assume a fixed N/O ratio as a function of metallicity. This is a limitation of the \sphinx{} data as elements are not enriched individually and we thus do not capture the expected primary to secondary sequence as a function of metallicity \citep[e.g.][]{Pilyugin2012}. Decreasing N/O at lower metallicity would cause some galaxies to scatter left and impact our upper sequence. Similarly if C/O also decreases  at lower O/H, the number of cooling channels is reduced which would increase [O~{\small III}]/H$\beta$. We also note a significant number of galaxies fall above the classic \cite{Kewley2001,Kauffmann2003} demarcations between star-forming galaxies and AGN. This is consistent with high-redshift JWST observations \citep{Sanders2023,Cameron2023} and is a result of \sphinx{} galaxies achieving very high ionization parameters in extremely dense gas\footnote{Part of this is a modelling choice because in gas cells with unresolved Stromgren spheres, we act as if all constituent star particles are co-located rather than assuming they are uniformly scattered throughout the gas cell. In the latter case, the ionization parameter at the inner radius of the shell is significantly decreased.}. Since galaxies in \sphinx{} were selected to be star-forming, we do not sample the downturn in the BPT seen for low-redshift, much more passive galaxies with low-ionization parameters and high-metallicities. 

\spdrone\ is not limited to only the BPT diagram. In the bottom panel of Figure~\ref{fig:bpt} we show an example R23-O32 diagram for \sphinx{} galaxies compared to SDSS and $z>5$ JWST galaxies. Similar to the BPT, we find that \sphinx{} galaxies populate the higher-excitation and lower-metallicity regions of the diagram, coincident with high-redshift JWST galaxies. Because this diagram only depends on oxygen emission lines, it suffers fewer systematics in modelling compared to the BPT.

\subsubsection{Ly$\alpha$ luminosity function}
Similar to the nebular line luminosity functions, direct comparisons can be made between \sphinx{} and observations for Ly$\alpha$. For all redshifts we provide data to make intrinsic Ly$\alpha$ luminosity functions and for $z=4.64,\ 5, \&\ 6$ we provide Ly$\alpha$ luminosities for ten sight lines that have been post-processed with Monte Carlo radiation transfer out to the virial radius of the halo. We emphasize that IGM absorption can be significant at these redshifts \citep[e.g.][]{Inoue2014} and our modelling choice is such that additional IGM attenuation can be applied in post-processing to understand how the IGM neutral fraction impacts the observability of Ly$\alpha$ after it escapes the ISM and CGM.

Previous Ly$\alpha$ luminosity functions for \sphinx{} were presented in \cite{Garel2021} and \cite{Katz2022-mgii}. The differences between those works and here are that the former only studied the angle-averaged escape rather than the line-of-sight values that are part of this data release. Furthermore, this work has updated the method for estimating intrinsic Ly$\alpha$ emission, most notably the fix for unresolved Stromgren spheres. As this fix primarily occurs in very dense regions where Ly$\alpha$ photons are readily absorbed, our results are very similar to our earlier work.

\begin{figure}
\includegraphics[width=0.45\textwidth]{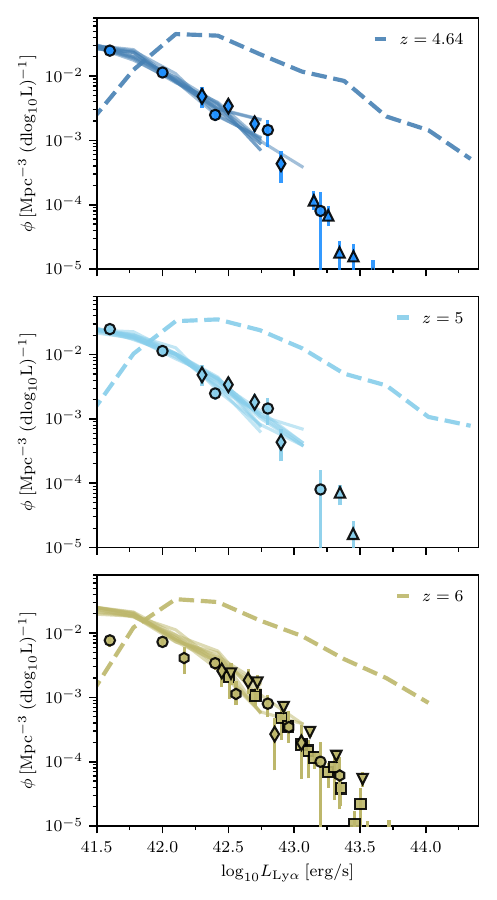}
\caption{Ly$\alpha$ luminosity functions for \sphinx{} galaxies at $z=4.64, 5$ and 6. Dashed lines show the intrinsic luminosity function while solid lines correspond to luminosity functions after radiation transfer (i.e. absorption + scattering, up to the virial radius of each halo) for ten viewing angles.  Circles, diamonds, upward triangles, downward triangles, hexagons, and squares show observational constraints from \protect\cite{Drake2017,Herenz2019,Sobral2018,Santos2016,Cassata2011,Konno2018}, respectively. Note that IGM transmission is not included in these figures.}
\label{fig:lya-lf_tg}
\end{figure}

In Figure~\ref{fig:lya-lf_tg} we show Ly$\alpha$ luminosity functions for \sphinx{} galaxies compared to observational constraints from $z=4.64-6$. Very good agreement is found between simulations and observations. The only discrepancy occurs at $z\sim6$ at $L_{\rm Ly\alpha}<10^{42}\ {\rm erg\ s^{-1}}$ where \sphinx{} predicts many more faint emitters than are observed. This may simply be due to flux limits, incompleteness corrections in the observations, or IGM attenuation (which is not included in our analysis). Nevertheless, we find very little evolution in the Ly$\alpha$ luminosity function of star-forming \sphinx{} galaxies in the redshift interval $z=4.64-6$. Future data releases may contain processed Ly$\alpha$ at higher redshifts as it has now been detected in very UV faint galaxies \citep{Saxena2023} up to $z\sim11$ \citep{Bunker2023}.

\subsubsection{Ly$\alpha$ and H$\alpha$ spectra and radial profiles}
In addition to Ly$\alpha$ luminosities and escape fractions, the \sphinx{} data release contains the full spectrum and radial profiles of Ly$\alpha$ and H$\alpha$. This allows for determining how the shape of the spectrum and its offset from line-centre relates to the intrinsic properties of the galaxy. 

\begin{figure}
\includegraphics[width=0.45\textwidth]{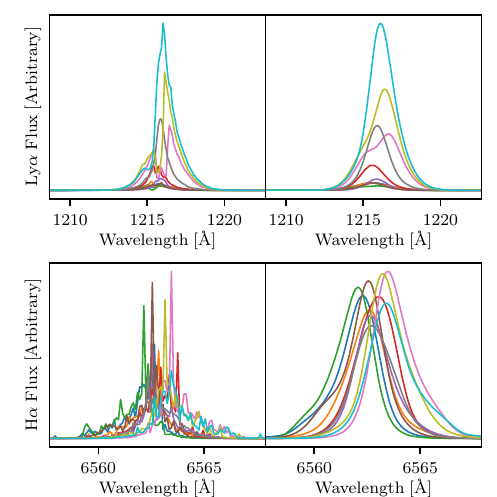}
\caption{(Left) Ly$\alpha$ and H$\alpha$ spectra along ten sight lines are shown for an example $z=6$ galaxy at the fiducial spectral resolution. (Right) The same spectra after being convolved with a Gaussian filter representing an arbitrary line-spread function.}
\label{fig:lya-ha-spec}
\end{figure}

In Figure~\ref{fig:lya-ha-spec} we show the Ly$\alpha$ and H$\alpha$ spectrum of an example $z=6$ galaxy. The top left panel demonstrates the diversity of spectral profiles one can see for Ly$\alpha$ for the same galaxy \citep[e.g.][]{Blaizot2023}. The spectrum varies between complete line-centre emission and full absorption. The H$\alpha$ spectrum exhibits less variance in terms of luminosity; however, there are strong features in the spectrum that correspond to individual star-forming regions in the galaxy. Spectra, especially those at high-redshift, are rarely sampled as well as in the \sphinx{} data release and are subject to a line-spread function. To demonstrate the impact of the line-spread function, we convolved the Ly$\alpha$ and H$\alpha$ spectra using a 1D Gaussian filter across five spectral pixels and the results are shown in the right panels of Figure~\ref{fig:lya-ha-spec}. Where previously the H$\alpha$ spectra exhibited a significant number of features, the convolved spectra appear significantly more Gaussian as one typically sees with a real spectrograph. Similarly, the features of the Ly$\alpha$ spectra are less well resolved after the convolution. We note that the spectra have been computed in the rest-frame of the simulation box rather than that of the galaxy. Hence when viewing the galaxy along different sight lines, the peculiar velocities can shift the emission line from line-centre. This can be corrected (as an observer might do) by shifting the spectrum in velocity-space so that H$\alpha$ falls on line-centre. Finally, we re-emphasize that Ly$\alpha$ spectra have only been integrated to the virial radius of the galaxy and not through the IGM. Hence further IGM attenuation must be applied to understand how these galaxies may appear in regions with different neutral fractions. 

\begin{figure}
\includegraphics[width=0.45\textwidth]{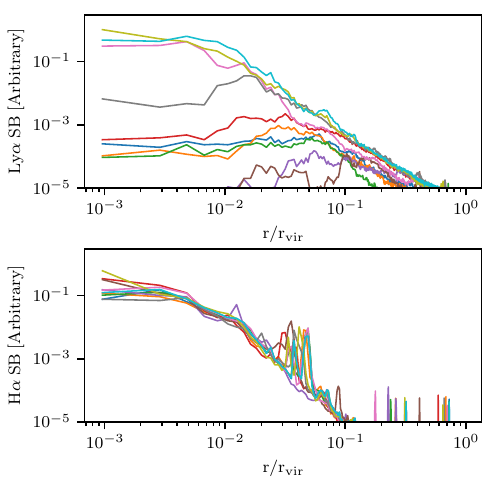}
\caption{Ly$\alpha$ and H$\alpha$ radial surface brightness profiles for the same galaxy as shown in Figure~\protect\ref{fig:lya-ha-spec}.}
\label{fig:lya-ha-sb}
\end{figure}

In Figure~\ref{fig:lya-ha-sb} we show the Ly$\alpha$ and H$\alpha$ radial profiles of the same galaxy. For this particular object, H$\alpha$ is much more centrally concentrated than Ly$\alpha$ as is does not resonantly scatter off of H~{\small I} and D. For many sight lines, there is no Ly$\alpha$ emission at all in the central regions consistent with the spectral profiles that show complete absorption. The comparison between the two profiles could provide indications into the state of the ISM and CGM at high redshift and the LyC escape fraction (Choustikov~et~al. {\it in~prep.}).

\section{Discussion}

\subsection{Comparison with Other Simulations}
Although the \sphinx{} data release is one the most complete data sets in terms of breadth of observational comparisons that can be made, our work is by no means the first simulated data set for which mock observations have been computed. In this section, we contextualize the \sphinx{} data release with respect to other simulations where mock observations were published by briefly highlighting some of the similarities and differences.

\subsubsection{Comparison with semi-analytic models}
Semi-analytic models are a computationally efficient way of producing mock observational catalogs for a large population of galaxies. Simplistic models are often applied to dark matter halo merger-trees in order to predict the physical properties of galaxies. In the context of JWST, numerous such models have been published \citep[e.g.][]{Williams2018,Yung2019}. The benefit to such models is that they can be tuned to match a wide range of galaxy properties and predict the full SEDs of galaxies as they evolve. Such models have been readily applied to design JWST observations in early cycles when little was known about the spectroscopic properties of the high-redshift galaxy population. However, due to their simplicity, semi-analytic models lack detailed spatial information of galaxy properties, they cannot self-consistently predict the evolving ISM properties of galaxies which are crucial for emission line predictions and dust attenuation, and finally, they often tie star formation with mass accretion, which conflicts with the bursty star formation histories often predicted in cosmological simulations at early epochs \citep[e.g.][]{Ma2018}. 

Due to computational expense, \sphinx{} only contains a single set of subgrid models and is thus not as ideal for sampling the wide parameter space of physical recipes in the way that semi-analytic models provide. However, \sphinx{} is a significant improvement over such models as it resolves the vast majority of physics that is important for generating the emission that is observed in a cosmological volume large enough to make meaningful comparisons with observations. 

\subsubsection{Comparison with hydrodynamic simulations}
Over the past decade, large suites of cosmological hydrodynamics simulations \citep[e.g.][]{Vogelsberger2014,Dubois2014,Schaye2015} have been run that can reproduce many of the observational properties and the diversity of galaxies at low-redshift. Because the subgrid models are tuned to reproduce the low-redshift galaxy population, if is often the case that the models differ at high-redshift. Thus JWST, observations are a crucial ingredient for differentiating between galaxy formation models. In general, these hydrodynamic simulations have much larger volumes compared to \sphinx{} due to the computational demands of on-the-fly radiative transfer. Therefore, they capture the brighter end of the UV luminosity function in a regime where our current work cannot make predictions. Hence they represent a complementary view of the high-redshift Universe.

\cite{Vogelsberger2020} recently presented UV and H$\alpha$ luminosity functions for multiple dust models for galaxies in the {\small IllustrisTNG} simulations. In comparison to our work, due to the much higher spatial resolution and mass resolution of \sphinx{} compared to {\small IllustrisTNG}, as well as ISM modelling, we find burstier star formation histories. This results in an increased scatter in the relation between UV magnitude and stellar mass. For example, at $M_{\rm UV}=-18$, we find a scatter of $\sim0.5$~dex in stellar mass compared to \cite{Vogelsberger2020} where the value is typically $\lesssim0.2$~dex. Our value is in much better agreement with observations \citep{Song2016}, even if the normalization of \sphinx{} is offset in this relation. Because {\small IllustrisTNG} does not model the ISM, H$\alpha$ luminosities are painted on to star particles in order to predict an H$\alpha$ luminosity function. Hence their predictions are not properly modulated by the different ISM conditions (e.g. gas temperature, density, metallicity contrast between dust and stars, LyC escape fraction, diffuse ionized gas, etc., see \citealt{Tacchella2022}) that impact H$\alpha$ emission. Similar to the UV luminosity function, their model will predict significantly less scatter than ours for a given star formation rate. 

\cite{Hirschmann2022} have performed an alternative post-processing of {\small IllustrisTNG} compared to \cite{Vogelsberger2020} in order to predict the evolution of emission line luminosities and diagnostic diagrams. The upside to these models compared to \sphinx{} is that they sample both star-forming galaxies and AGN and they have an explicit model for capturing emission from shocks when it is not always well resolved in \sphinx{}. However, the major limitation still remains that {\small IllustrisTNG} does not model the ISM. For this reason, \cite{Hirschmann2022} must assume a gas density (which they choose to be $10^2~{\rm cm^{-3}}$) in H~{\small II} regions; whereas observations show that this value likely increases with redshift \citep{Isobe2023}. Despite \sphinx{} having a much smaller volume compared to {\small IllustrisTNG-50} (20~cMpc vs. 50~cMpc), we find that our emission line luminosities tend to be higher than those in \cite{Hirschmann2022}. For example, the brightest [O~{\small III}]~$\lambda$5007 emitter at $z=6$ in {\small IllustrisTNG-50} (based on their Figure~11) has a similar intrinsic luminosity to the one in \sphinx{}. This once again may be due to the burstier nature of star formation in \sphinx{} compared to {\small IllustrisTNG} as well as less regulatory feedback in \sphinx{}. Other factors that could impact this result are differences in ISM gas density (\sphinx{} has much higher gas densities which can lead to higher intrinsic emission) or mass-metallicity relation. 

The post-processing technique also matters significantly in terms of mock observations. \cite{Shen2020} also post-processed {\small IllustrisTNG} and find significantly different emission line luminosities compared to \cite{Hirschmann2022}. Hence when the ISM is not resolved, the freedom in post-processing parameters likely represents the dominant systematic uncertainty in the ability to make observational predictions with simulations.

{\small BlueTides} \citep{Wilkins2016,Wilkins2017,Marshall2022} is one of the largest full-box cosmological simulations of high-redshift galaxy formation. Due to its large volume, it probes galaxies orders of magnitude more massive than what is in \sphinx{} and hence it is much better for understanding the physics dictating the bright-end of the galaxy UV luminosity function. Like {\small IllustrisTNG}, {\small BlueTides} does not have an ISM so even though the full SEDs of galaxies can be predicted, the impact of nebular emission is calculated by assigning photoionization models with a density of $10^2~{\rm cm^{-3}}$ and an escape fraction of zero to star particles. This ignores any potential difference between stellar and nebular metallicity (as we show above can be large in the case of a very dense ISM) and essentially fixes the ionization parameter for a given stellar age and metallicity. Similar to \sphinx{}, \cite{Wilkins2016} find that the nebular continuum can be an important component for high-redshift galaxy SEDs; however, the mass ranges of the two simulations do not overlap well enough to be compared. 

The {\small FLARES} simulations \citep{Lovell2021} are a suite of high-redshift zoom simulations employing the {\small EAGLE} model \citep{Crain2015} for galaxy formation. {\small FLARES} publicly provides numerous galaxy properties as well as SED fluxes, line luminosities, and EWs, similar to the \sphinx{} data release. While not a full volume simulation, the suite is large enough to contain a diversity of galaxies. For example, {\small FLARES} contains high-redshift galaxies with Balmer breaks \citep{Wilkins2023} that are also present in \sphinx{} (see \citealt{Steinhardt2023}) and have been observed at high redshift \citep[][]{Hashimoto2018}, although c.f. \cite{Bradac2023,Stiavelli2023}. Similar to {\small BlueTides} and {\small IllustrisTNG}, {\small FLARES} does not have an ISM so nebular emission is assumed to originate from an ionization-bounded nebula with a maximum ionization parameter of $-2$, assuming that gas near the star particles have the same metallicity as the star. One of the notable differences between \sphinx{} and {\small FLARES} is that many of our galaxies can reach [O~{\small III}]~$\lambda$5007/H$\beta>10$ which are not seen in {\small FLARES}. This may be due to the fact that in general \sphinx{} galaxies can exhibit ionization parameters much greater than $-2$. However, {\small FLARES} contains a higher proportion of galaxies with [O~{\small III}]~$\lambda$4959,5007$+$H$\beta>1000~$~\angstrom{} in better agreement with observations. This may be due to differences in dust modelling and star formation recipe.

{\small FirstLight} \citep{Ceverino2019} adopts a similar approach to {\small FLARES} by using multiple zoom simulations to capture the diversity of galaxies at high-redshift. Unlike the previously discussed simulations in this section, {\small FirstLight} attempts to self-consistently model the ISM. Despite this, nebular regions are modelled using a constant gas density of $10^2~{\rm cm^{-3}}$ and intrinsic emission line luminosities are adopted from those provided with \bpass{} \citep{Xiao2018}. 

{\small FirstLight} deals only with intrinsic emission rather than dust-attenuated quantities. Galaxies in {\small FirstLight} exhibit bursty star formation histories, similar to \sphinx{} and they similarly find that the nebular continuum sets a lower limit on UV slope (although ours is $\sim-2.7$ whereas it is closer to $-2.5$ in their work). One considerable difference is the lack of scatter in the N~{\small II} BPT diagram in {\small FirstLight} compared to \sphinx{} as well as local metal poor galaxies \citep{Guseva2017}. Furthermore, \sphinx{} probes [O~{\small III}] emitters with observed, rest-frame equivalent widths $>1000 \ \text{\AA} $ that are not absent from the intrinsic estimates in {\small FirstLight}. This is partially due to the fact that \sphinx{} probes galaxies with higher sSFR; but, also possibly due to the high densities reached in the nebular regions in \sphinx{}.

\subsubsection{Comparison with radiative-hydrodynamic simulations without a multiphase ISM}
\label{sec:rt_comp_no_ism}
Simulations like {\small THESAN} \citep{Kannan2022} improve upon their related predecessor {\small IllustrisTNG} due to the addition of on-the-fly radiation hydrodynamics. Public data is provided for Ly$\alpha$, SEDs, and various other galaxy properties similar to {\small FLARES} and now \sphinx{}. The volume of {\small THESAN} is significantly larger than \sphinx{}; hence, it probes much more massive galaxies and a more diverse IGM. Moreover, {\small THESAN} includes black holes which are absent from \sphinx{} and may be relevant at high redshift \citep[e.g.][]{Greene2023,Matthee2023b,Kocevski2023}.

Like the simulations in the previous section, {\small THESAN} does not have a multiphase ISM which means that Ly$\alpha$ and nebular predictions are entirely subject to the ISM assumptions made in post-processing. The same holds true for the LyC escape fraction estimates from {\small THESAN}. As with {\small IllustrisTNG} this can lead to order of magnitude systematic uncertainties for emission predictions from the same simulation. This is particularly true for the intensity-mapping predictions presented in \cite{Kannan2022a} regarding [C~{\small II}]~158$\mu$m emission, which is predicted to mostly come from neutral ISM gas at high-redshift \citep[e.g.][]{Pallottini2017}, not H~{\small II} regions with a fixed ionization parameter of $-2$. We have processed \sphinx{} with sub-ionizing radiation specifically to better capture the physics of photodissociation regions. For similar reasons, without an ISM, {\small THESAN} cannot reliably predict the spectral shapes of Ly$\alpha$ escaping the ISM \citep{Smith2022} which are set by the complex kinematics and ionization state of the dense gas near H~{\small II} regions \citep[e.g.][]{Blaizot2023}. The simulation is better suited than \sphinx{} for measuring IGM transmission curves, similar to {\small CODA} \citep{Park2022} due to the large volume compared to \sphinx{}. 

\begin{figure}
\includegraphics[width=0.45\textwidth]{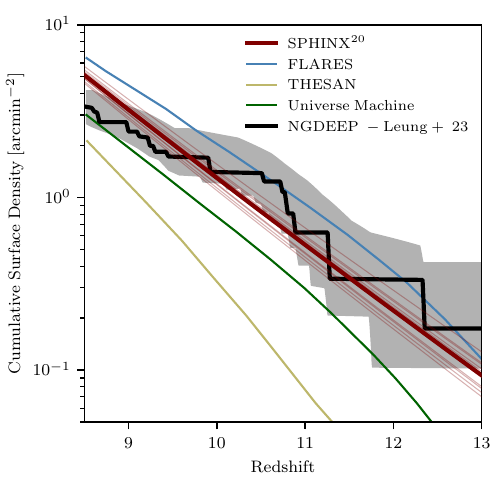}
\caption{Cumulative surface density of sources with F277W magnitudes $<29.5$ at redshifts $>z$ for \sphinx{} galaxies (red) compared to JWST NGDEEP observations \citep[black,][]{Leung2023} and other simulations/models. The thick red line shows the mean across all sight lines in our simulations while the thin red lines represent individual viewing angles. The grey shaded region represents the uncertainty on the observations.}
\label{fig:cum_surface_density}
\end{figure}

One final notable difference is that \sphinx{} exhibits a much higher conversion efficiency of baryons into stars compared to {\small THESAN}. While the high efficiency is less in agreement with pre-JWST abundance matching estimates \citep[e.g.][]{Behroozi2019}, and may still be too efficient given our discussion comparing JADES and \sphinx{} photometry, these older estimates are inconsistent with the surface densities of bright galaxies at high-redshift \citep[e.g.][]{Finkelstein2023,Leung2023}. This is shown in Figure~\ref{fig:cum_surface_density} where we compare the cumulative surface density of sources with F277W magnitudes $<29.5$ from \sphinx{} to recent results from NGDEEP \citep[e.g.][]{Leung2023} and other simulations\footnote{All curves for this plot besides those from \sphinx{} were digitized \citep{Rohatgi2022} from Figure~4 in \cite{Leung2023}.}. Indeed {\small THESAN} and {\small Universe Machine} \citep{Behroozi2019} under predict the results from NGDEEP while \sphinx{} is in agreement with the data up to $z\sim9$, where it then begins to over predict the galaxy counts. 

\subsubsection{Comparison with zoom-in radiative-hydrodynamic simulations}

The {\small Renaissance} simulation \citep{Oshea2015} is a suite of high resolution cosmological radiation hydrodynamics zoom simulations that samples large regions from under-dense, to over-dense, stopping at progressively higher redshift. The spatial resolution is higher than \sphinx{} and the stellar masses in {\small Renaissance} are in good agreement with inferences from high-redshift JWST observations \citep{McCaffrey2023}. SEDs and JWST photometry for {\small Renaissance} galaxies were presented in \cite{Barrow2017} and due to the simulation volume, {\small Renaissance} samples much lower mass galaxies, and only at higher redshifts. The post-processing method shares many similarities with the methods presented in this work. Consistent with \sphinx{}, they find a large scatter in the [N~{\small II}] BPT diagram at low-metallicity and that scattering can play an important role in modulating the observed spectrum. Similar analysis techniques were applied to cosmological radiative hydrodynamics simulations of \cite{Barrow2020} who find similar results to \sphinx{} (specifically \citealt{Choustikov2023}) with regards to how LyC escape trends with nebular emission.

Numerous other zoom-in cosmological radiation hydrodynamics simulation of individual high-redshift objects have been presented in the literature \citep[e.g.][]{Katz2017,Katz2019,Pallottini2017,Pallottini2019,Lupi2020,Trebitsch2021}. For those that provide emission lines and spectra, post-processing techniques are similar to those used in this work --- i.e. non-equilibrium emission is computed when possible, otherwise photoionization models are used. The simulations of \cite{Katz2019} are most similar to \sphinx{} in terms of mass, spatial resolution, and feedback, although that work included more radiation bins (primarily sampling lower energies), the dust model followed \cite{RR2014}, and molecular hydrogen was self-consistently modelled. No significant differences can be reported in terms of emission line properties, validating the choice of a metallicity floor and the RT post-processing in the context of the emission lines studied here.

\subsection{Future Updates}
\label{sec:future}
While the \sphinx{} data release represents the current state-of-the-art in comparing high-redshift galaxy simulations with observations, we consider the data set to be an evolving tool rather than a static reference. We describe below some of the intended upgrades following the initial data release.

\begin{enumerate}
    \item {\bf Larger galaxy sample}: The current data release contains only $\sim 1,400$ galaxies at seven different redshift bins. This is significantly fewer than the nearly 30,000 resolved galaxies in each \sphinx{} snapshot. Data is stored from the simulation on a 5~Myr cadence; hence, there is significant room to increase the sample size of the data release. This may help capture the even rarer, extreme objects that represent the ``tip-of-the-iceberg'' of what is currently being observed with the first JWST observations. The limiting factors in providing all galaxies in the initial release are due to computation time and data storage and these will be less problematic moving forward.
    \item {\bf Extensions beyond star-forming galaxies}: A star-formation threshold of $0.3~{\rm M_{\odot}~yr^{-1}}$ was chosen to limit the galaxies that were being post-processed to only those that are likely observable with JWST. However, JWST observes flux, not SFR. Without post-processing the data, we are unable to know which galaxies have high enough magnitudes to be observed; hence SFR was our proxy. Unfortunately this choice means that certain classes of galaxies, e.g. quiescent or remnant leakers, are absent from the initial data release, despite the fact that they may represent critical phases in galaxy evolution. Future releases would ideally contain \sphinx{} galaxies that are in between star formation events rather than at a peak.  
    \item {\bf Alternative dust models}: Since \sphinx{} does not explicitly follow the formation and destruction of dust, we have adopted an effective dust model that assumes a particular SMC attenuation curve and connects the dust content with the neutral gas. While this model is successful in that our UV luminosity and Ly$\alpha$ luminosity functions are in good agreement with observations, it is not clear whether this model is unique in that regard. For this reason, we plan to sample more dust models to better understand their effect on mock observations. 
    \item {\bf Additional emission lines}: The data set currently contains the vast majority of emission lines that have been observed by JWST at high-redshift; however, there are notable exceptions (for example Mg~{\small II}~$\lambda$2796,2803 and N~{\small IV}~$\lambda$1483,1486). We have also not provided most of the IR lines that can be observed with ALMA. Such additional emission lines may be part of future data releases.
    \item {\bf Varying metal abundance patterns}: Because \sphinx{} does not follow enrichment of individual elements, we have scaled solar abundance patterns to the metallicity of each gas cell. However, it is well known that ratios such as C/O and N/O vary with metallicity. This effect is important for both line ratios and cooling in the ISM and future data releases may attempt to address this effect.
    \item {\bf Spectral data cubes}: While the current data release contains line luminosities of the integrated galaxy spectra, which is comparable to the vast majority of JWST spectral data, one of the unique aspects of JWST is its IFU capabilities. {\small RASCAS} has the capabilities of outputting spectral data cubes for any emission line and the limiting factor in their public release is purely data storage. These can be currently made available on-request for individual galaxies and will ideally be standard in future releases.
    \item {\bf Ly$\alpha$ IGM propagation}: The Ly$\alpha$ spectra that we released are calculated only out to the virial radius of the galaxy. While this is instructive for understanding how the ISM and CGM modulate the intrinsic Ly$\alpha$ emission, it is not directly comparable to high-redshift galaxy observations where the CGM remains important, even in the post-reionization epoch \citep[e.g.][]{Inoue2014}. Although the \sphinx{} volume does not quite probe the cosmological homogeneity scale ($\sim100$~Mpc), we are still able to assess the impact of the IGM on our spectra more locally (see e.g. \citealt{Garel2021}).
    \item {\bf Environmental Information}: Additional environmental information, such as neighbour properties, filamentary properties, and whether the galaxies reside in ionized bubbles may also be useful for comparing with observations and understanding the role of radiation feedback on galaxy properties (see e.g. \citealt{Katz2020reion}). For example, the recent observation that high-redshift Ly$\alpha$ emitters tend to have companions \citep{Witten2023b} would not currently be testable with the \sphinx{} data set as we only include positions of star-forming galaxies. 
\end{enumerate}

\subsection{Caveats}
The \sphinx{} data release represents a unique tool to both compare theory with observation and to help interpret observations; however, like all numerical simulations, there remain important limitations of the data set. Most importantly, \sphinx{} has finite spatial and mass resolution limiting how much of the ISM turbulent structure is truly resolved. While the resolution of \sphinx{} is the state-of-the-art for radiative hydrodynamics simulations of this size, in many cases, Stromgren spheres remain unresolved. This is in many ways a result of the way stars form in the simulation (see discussion in \citealt{Rosdahl2015}), but the modelling choices have a strong impact on our results. For example, in the case where multiple star particles exist in a cell where the net Stromgren sphere is unresolved, we model this as a single source at the centre of the cell. Thus the ionization parameter is much higher than the case where we model these as separate H~{\small II} regions embedded in the same cell. Choosing the former can result in higher e.g. [O~{\small III}]/H$\beta$ than the latter method. Finite spatial resolution also has a strong impact on Ly$\alpha$ and LyC escape physics. It is not yet known how well these quantities are converged with resolution. Furthermore, due to the fixed mass and comoving spatial resolution, high-mass galaxies are better resolved than low mass ones at a fixed redshift and the ISM is better resolved in the higher redshift snapshots compared to the lower redshift ones. The important impact of this numerical effect is that the gas can reach higher densities in more massive galaxies and at high-redshift for a fixed halo mass. The gas density is a fundamental parameter that impacts star formation and much of the emission presented in this work. Intrinsic emission will be more strongly affected as in nearly all galaxies, the resolution in \sphinx{} is high enough such that the densities impacted by this effect are typically optically thick (at short wavelengths).

One may argue that due to the inevitable use of subgrid models and the imperfect match between simulations and all high-redshift observations, \sphinx{} is not a completely accurate representation of high-redshift galaxy formation. This is not unique to \sphinx{} and likely true of all galaxy formation simulations. However, photoionization models that are very simplistic have been the dominant means of interpreting high-redshift spectra. {\bf Despite the fact that \sphinx{} is imperfect, it undoubtedly captures a much more complex ISM and SFHs that better reflect reality than photoionization models, especially since a \sphinx{} galaxy can be thought of as a complex combination of photoionization models with the additional elements of diffuse gas and a non-zero escape fraction.} This is consistent with the fact that groups are now adopting combinations of photoionization models to explain emission lines rather than relying on individual models \citep[e.g.][]{Ramambason2022}. Moreover, \sphinx{} provides a unique test-bed for understanding how well ISM properties can be inferred from an integrated SED. In this case, it does not matter in which galaxy the ISM resides, this is simply a photon counting exercise for a given density, metallicity, and stellar population distribution. For this reason we argue that the \sphinx{} data release represents a drastic improvement over previous methods. At worst, our results should be interpreted as more complex, 3D photoionziation models. This will only be superseded by simulations where non-equilibrium metal chemistry is fully coupled to radiative transfer \citep[e.g.][]{Katz2022-rtz} and the Stromgren spheres are resolved \citep[e.g.][]{Kimm2022}.

In addition to subgrid models, \sphinx{} also relies on a series of modelling assumptions. For example, we have chosen a specific stellar IMF (Kroupa-like), a stellar population synthesis library ({\small BPASS}) for the stellar spectra, a particular dust model (SMC), etc. Different choices for these parameters can systematically shift galaxy properties such as mass-to-light ratios, ionizing photon luminosities and therefore emission line luminosities, dust attenuation as a function of wavelength and metallictiy and thus ${\rm E(B-V)}$ values and $\beta$ slopes. Once again, this is not only true for \sphinx{}, but is also true for other numerical simulation as well as SED fitting codes. For these reasons, we have tried to compare directly with observations when possible rather than derived or inferred quantities to mitigate these systematic biases between codes. 

\section{Conclusions}
Here we have described Version 1 of the \sphinx{} data release, which is publicly available to download from \url{https://github.com/HarleyKatz/SPHINX-20-data}. \sphinx{} is currently the largest full box simulation of galaxy formation in the epoch of reionization with a multiphase ISM and thus represents a valuable resource to the wider community, especially given the recent launch of JWST. The data release contains galaxies spanning $\sim5$~dex in stellar mass and star formation rate that exhibit a diverse set of mass growth and star formation histories. The unique aspect of this data set is the focus on forward modelling observational quantities such that the data can be directly compared with observations. We provide full spectra and photometry including self-consistently calculated nebular emission lines (including Ly$\alpha$) and continuum, accounting for absorption and scattering by dust (and H~{\small I}). We have provided a tour of all of the different quantities available in the data set and made comparisons with ERS and Cycle~1 JWST data where possible to demonstrate the utility of the data in interpreting high-redshift observations and to show where constraints can be made on the physics of early galaxy formation.

While the results presented here represent a static view of the \sphinx{} public data release, we emphasize, as indicated in Section~\ref{sec:future} that the data set will be updated continuously over time to provide both new data and enhance existing data products. We encourage users to quote the git hash of the data set when publishing with \sphinx{} data for reproducibility purposes. 

The data is organized in a series of CSV tables as well as {\small JSON} files that have been indexed by halo ID and redshift. These files can be flexibly loaded into any relevant computing software (e.g. Python, IDL, Julia, etc.). In addition to the curated data, we also release a repository of Juptyer notebooks with basic tutorials on how to load and manipulate the data in Python. Our aim is to lower the barrier to entry for interested users of high-redshift galaxy formation simulations and to ease the comparison between simulations and observations.

\section*{Acknowledgements}

HK thanks Jonathan Patterson for support with the local Glamdring cluster as well as Adrianne Slyz and Julien Devriendt for access to dp016 computing hours on DiRAC.  TK was supported by the National Research Foundation of Korea (NRF-2022R1A6A1A03053472 and NRF-2022M3K3A1097100).

This work used the DiRAC@Durham facility managed by the Institute for Computational Cosmology on behalf of the STFC DiRAC HPC Facility (\url{www.dirac.ac.uk}). The equipment was funded by BEIS capital funding via STFC capital grants {\small ST/P002293/1}, {\small ST/R002371/1} and {\small ST/S002502/1}, Durham University and STFC operations grant {\small ST/R000832/1}. DiRAC is part of the National e-Infrastructure. This work was performed using the DiRAC Data Intensive service at Leicester, operated by the University of Leicester IT Services, which forms part of the STFC DiRAC HPC Facility (\url{www.dirac.ac.uk}). The equipment was funded by BEIS capital funding via STFC capital grants {\small ST/K000373/1} and {\small ST/R002363/1} and STFC DiRAC Operations grant {\small ST/R001014/1}. DiRAC is part of the National e-Infrastructure.

Computing time for the SPHINX project was provided by the Partnership for Advanced Computing in Europe (PRACE) as part of the ``First luminous objects and reionization with SPHINX (cont.)'' (2016153539, 2018184362, 2019215124) project. We thank Philipp Otte and Filipe Guimaraes for helpful support throughout the project and for the extra storage they provided us. We also thank GENCI for providing additional computing resources under GENCI grant A0070410560. Resources for preparations, tests, and storage were also provided by the Common Computing Facility (CCF) of the LABEX Lyon Institute of Origins (ANR-10-LABX-0066) and PSMN (Pôle Scientifique de Modélisation Numérique) at ENS de Lyon.

The main roles of the authors were, using the CRediT (Contribution Roles Taxonomy) system\footnote{\url{https://authorservices.wiley.com/author-resources/Journal-Authors/open-access/credit.html}}:
\textbf{Harley Katz:} conceptualization; methodology, investigation, data curation, writing -- original draft; formal analysis; visualization; software, 
\textbf{Joki Rosdahl:} conceptualization; writing -- original draft; formal analysis; software; validation; visualization, 
\textbf{Taysun Kimm:} writing -- original draft; formal analysis; validation; visualization, 
\textbf{Jeremy Blaizot:} writing -- original draft; formal analysis; validation; visualization, 
\textbf{Nicholas Choustikov:} validation; software;  writing -- review \& editing, 
\textbf{Marion Farcy:} writing -- original draft; formal analysis; validation; visualization, 
\textbf{Thibault Garel:} writing -- original draft; formal analysis; validation; visualization, 
\textbf{Martin Haehnelt:}  writing -- review and editing, 
\textbf{Leo Michel-Dansac:} software, writing -- review and editing 
\textbf{Pierre Ocvirk:} writing -- original draft; formal analysis; validation; visualization.

\bibliographystyle{mn2e}
\bibliography{library_oj}

\end{document}